\documentclass[twocolumn,preprintnumbers,superscriptaddress]{revtex4}
\usepackage[dvipsnames]{xcolor}
\usepackage{fancyhdr}
\usepackage{hyperref}
\usepackage{physics}
\usepackage{graphicx}
\usepackage{amsmath}
\usepackage{amssymb}
\usepackage{qcircuit}
\usepackage{xspace,slashed}
\usepackage[OT1]{fontenc}
\usepackage{cmbright}
\DeclareFontShape{OT1}{cmss}{m}{it}{<->ssub*cmss/m/sl}{}

\usepackage{xcolor}
\usepackage{braket}
\usepackage{bm}

\newcommand{\Qibolab}{\texttt{Qibolab}\xspace}
\newcommand{\Qibo}{\texttt{Qibo}\xspace}
\newcommand{\Qibocal}{\texttt{Qibocal}\xspace}
\newcommand{\Qibosoq}{\texttt{Qibosoq}\xspace}

\fancypagestyle{firststyle}{
    \fancyhead[HR]{TIF-UNIMI-2023-13, CERN-TH-2023-157}
}

\begin{document}

\title{Multi-variable integration with a variational quantum circuit}

\newcommand{\MIaff}{TIF Lab, Dipartimento di Fisica, Universit\`a degli Studi
  di Milano and INFN Sezione di Milano, Milan, Italy.}
\newcommand{\TII}{Quantum Research Center, Technology Innovation Institute, Abu Dhabi, UAE.}
\newcommand{\CERNaff}{CERN, Theoretical Physics Department, CH-1211
  Geneva 23, Switzerland.}

\author{Juan M. Cruz-Martinez}
\affiliation{\CERNaff}
\email{juan.cruz.martinez@cern.ch}
\author{Matteo Robbiati}
\affiliation{\CERNaff}
\affiliation{\MIaff}
\author{Stefano Carrazza}
\affiliation{\CERNaff}
\affiliation{\MIaff}
\affiliation{\TII}

\begin{abstract}
In this work we present a novel strategy to evaluate multi-variable integrals with quantum 
circuits. The procedure first encodes the integration variables into a parametric
circuit.
The obtained circuit is then derived with respect to the integration variables
using the parameter shift rule technique.
The observable representing the derivative is then used as the predictor of the target 
integrand function following a quantum machine learning approach.
The integral is then estimated using the fundamental theorem of integral calculus by evaluating the original circuit.
Embedding data according to a reuploading
strategy, multi-dimensional variables can be easily encoded into the circuit's gates
and then individually taken as targets while deriving the circuit.
These techniques can be exploited to partially integrate a function or to
quickly compute parametric integrands within the training hyperspace. 
\end{abstract}

\maketitle
\thispagestyle{firststyle}

\section{Introduction}
\label{sec:introduction}

Many scientific and engineering problems require the evaluation of numerical integrals of varying complexity of the form:
\begin{equation}
  I(\bm{\alpha}) =  \int_{\bm{x}_a}^{\bm{x}_b} g(\bm{\alpha};\bm{x})\,\text{d}^n\bm{x},
  \label{eq:integral}
\end{equation}
where the bold symbols correspond to vectors, $\bm{x}_a$ and $\bm{x}_b$ define the integration domain and $g(\bm{\alpha};\bm{x})$ is an integrand which depends on the integration variables ($\bm{x}$) and some parameters ($\bm{\alpha}$).

There are many numerical integration methods which tackle this problem,
and the choice usually depends on the characteristics of the integrand functions and the integral region.
For instance, low-dimensional well-behaved integrals can be successfully integrated
with quadrature methods.
However, achieving the same precision with more complicated or higher-dimensional integrands will lead to a significant increase in computational costs.

In those cases, Monte Carlo (MC) methods are often favored due
to their ability to handle a wider range of integrand functions without imposing stringent requirements
and a convergence rate that does not depend on the dimensionality of the integrand.
An important feature of MC methods is the possibility of binning partial results so that
differential distributions in any of the integration variables can be obtained.
However, in exchange for their flexibility,
they suffer from slow convergence and require a large number of function evaluations.
To mitigate these issues, various techniques have been proposed to speed up the integration process, by reducing the number of integrand evaluations while producing accurate results~\cite{10.2307/2280232, caflisch_1998, zhong2022efficient, NIPS2002_24917db1,SCHMIDHUBER201585}.

In the context of particle physics, these advantages have made
the VEGAS~\cite{Lepage:1977sw,Lepage:2020tgj} MC algorithm into the gold standard for numerical integration.
Over the past decade, there have been numerous attempts to further improve the algorithm without modifying the underlying strategy.
Some examples are the implementation of the algorithm in new hardware devices~\cite{Carrazza:2020rdn,Gomez:2021czl}, which offer a raw speed-up over traditional computing;
the usage of multi-channel techniques~\cite{Kleiss:1994qy}, which exploit prior knowledge of the behavior of the different pieces of the integrand;
or machine learning techniques to enhance the importance sampling algorithm~\cite{10.1145/3341156,Bothmann:2020ywa}.

However, all these methods suffer from the same drawback:
obtaining a result requires the repeated numerical evaluation of the integrand in the region of interest.
In addition, even if the integral smoothly depends on parameters which are not integrated over (like $\bm{\alpha}$ in Eq.~\eqref{eq:integral}),
any change in them requires a complete new run.

In Refs.~\cite{DBLP:journals/corr/abs-2012-01714, Ma_tre_2023} a new approach has been proposed which can be utilized to circumvent this issue.
These methods are based on using artificial Neural Networks (aNN) to build a surrogate model for the primitive of the integrand.
Such a surrogate can keep the dependence on both the integration variables and the function parameters
within the integration domain in which it has been trained.
By subsequently training the derivative of the model to approximate the integrand it is possible to recover its primitive.
While the computational cost does not disappear (it is translated to the training process),
it allows for the evaluation of the integral (or the marginalization of the integrand over any of the variables)
for any choice of parameters within the training range at almost zero additional cost.

In this paper we apply the same strategy in a Quantum Machine Learning (QML)~\cite{Schuld_2014, 
Biamonte_2017, Mitarai_2018, chen2020variational, Abbas_2021} context
to estimate the value of integrals of the form of Eq.~\eqref{eq:integral}.

The unique aspects of quantum circuits renders them
an exceptional tool for this methodology.
In the classical version, we need to train the derivative of the aNN.
By taking the derivative, the architecture of the network can considerably change,
possibly leading to a costly hyperparameter search in order to find the optimal model~\cite{Ma_tre_2023}.
With a quantum circuit instead, we can exploit their properties in order to obtain the
derivative using the Parameter Shift Rule (PSR)~\cite{Schuld_2019,crooks2019gradients,Mari_2021,Wierichs_2022},
therefore using the same architecture for the derivative and its primitive.

For doing this, we use \texttt{Qibo}~\cite{Efthymiou_2021, Efthymiou_2022,
Carrazza_2023, pasquale2023opensource, stavros_efthymiou_2023_7736837, 
stavros_efthymiou_2023_7748527, andrea_pasquale_2023_7662185}, a full-stack and 
open-source framework for quantum simulation, control and calibration. 

The rapid development of quantum technologies leads us to believe that 
quantum circuits and QML tools can be exploited to improve on the performance,
despite the constraints imposed by the current era of limited Near-Intermediate Scale 
Quantum~\cite{Preskill_2018} (NISQ) devices.
In particular, we note quite some interest on the field of High Energy Physics
where many new algorithms are being developed and tested
leading to a very robust ecosystem of quantum computing tools focusing on
particle physics~\cite{delgado2022quantum, Gustafson:2022dsq, Agliardi:2022ghn, 
Bauer:2022hpo, wozniak2023quantum, Chawdhry:2023jks, robbiati2023determining, DElia:2024pax}

This paper is structured as follows, 
we expose the method in Sec.~\ref{sec:methodology}, after a brief introduction to QML and 
circuits derivative calculation respectively in Sec~\ref{subsec:qml} and 
Sec.~\ref{subsec:psr}. 
In Sec.~\ref{sec:results} we apply the method to two situations,
a toy-model represented by a $d$-dimensional trigonometric function
and a real-life scenario motivated by particle physics.

All results can be reproduced using the code at:

\href{https://github.com/qiboteam/QiNNtegrate}{\color{blue}  \texttt{https://github.com/qiboteam/QiNNtegrate}}.

\section{Methodology} 
\label{sec:methodology}

In this section we introduce well known concepts of quantum computation which are 
useful to better understand the methodology. Then, we describe the integration procedure 
more in detail.

\subsection{Quantum Machine Learning in a nutshell}
\label{subsec:qml}

Classical Machine Learning (ML) is nowadays widely used to tackle statistical
problems, such as classification, regression, density estimation, pattern recognition, etc.
The goal of the ML algorithms is to teach a model to perform some specific task 
through an iterative optimization process.
Let us recall here some basic ML concepts which equally apply to QML in order 
to establish the notation used through the paper. For simplicity, we treat here 
the case of Supervised Machine Learning, but what follows can be easily extended to
other approaches.

We consider two variables: an input vector $\bm{x}$ and an output 
vector $\bm{y}$, which is related to $\bm{x}$ through some hidden law
\begin{equation}
  \bm{y}=f(\bm{x}), \label{eq:basicML}
\end{equation}
which we aim to estimate.
These vectors can be composed of any number of variables and the dimensionality of the input and output data
can in general be different.
A model $\mathcal{M}_{\bm{\theta}}$, parametric in some parameters $\bm{\theta}$, is then chosen to make predictions 
$\bm{y}_{\rm est}(\bm{x}|\bm{\theta}) = \mathcal{M}_{\bm{\theta}}(\bm{x})$ of the output variables.
Once a model is selected, a loss function
$J$ is typically used to verify the predictive goodness of $\mathcal{M}_{\bm{\theta}}$.
Finally, an optimizer is required,
which is in charge of estimating:
\begin{equation}
\bm{\theta}_{\rm best} = \text{argmin}_{\bm{\theta}}  \left\{ \displaystyle\sum_{i} J\left[\bm{y}_{i,\rm est}(\bm{x}_i|\bm{\theta}), \bm{y}_{i,\rm meas}\right] \right\}. 
\end{equation}
where $\bm{y}_{\rm meas}$ is a measured realization of $\bm{y}$.

In Quantum Computation (QC) we use two-level quantum systems called \textit{qubits} to store information.
One of the most popular quantum computing paradigms is the \textit{Gate based Quantum Computing}, where  
the qubits state is manipulated with the action of unitary operators we call \textit{gates}, defining a 
totally reversible computation.
These unitaries can be combined to build up multi-qubit gates.
In practice, QC introduces new computational possibilities with respect to the classical counterpart with tools
such as superposition and entanglement.

In the context of Quantum Machine Learning, a common approach is to use Variational Quantum Circuits (VQC)~\cite{chen2020variational, Benedetti_2019, cerezo2020variational}
as a quantum version of the classical parametric models. 
A VQC is a collection of gates which depends on a set of parameters $\bm{\theta}$,
which can be used to regulate the manipulation of a quantum state.
A circuit can be applied to an initial quantum state $\ket{\psi_i}$ and, after the execution, 
measurements can be performed on the final state $\ket{\psi_f}$.
We can collect information by executing the circuit $N_{\rm nshots}$ times, and 
then calculating expected values of the target observables $\hat{O}$,
\begin{equation} 
    G_{\hat{O}}(\bm{\theta}) = \braket{ \psi_i| \mathcal{C}^{\dagger}(\bm{\theta}) \hat{O} \,
\mathcal{C}(\bm{\theta}) |\psi_i}.
\label{eq:qml_predictor}
\end{equation}    
In the following, for simplicity, we omit the reference to the observable, which
is chosen to be a non-interacting Pauli $\hat{\sigma}_z$ taken independently over all qubits and then averaged over the number of qubits.

The expectation values introduced in Eq.~\eqref{eq:qml_predictor} can be used as predictions 
$\bm{y}_{\rm est} = G(\bm{\theta})$ in the QML process. 

Several methods are known to embed data
into a quantum system~\cite{lloyd2020quantum, P_rez_Salinas_2020, incudini2022structure, Schuld:2018gao}.
In this work we follow the procedure presented in~\cite{P_rez_Salinas_2020} and known 
as \textit{data re-uploading}, which
allows us to encode multi-dimensional variables into the parametric gates of a circuit
as part of their parameters. Since we encode the data directly into the gates of 
the VQC, we don't need any additional state preparation and from now on we consider $\ket{\psi_i}=\ket{0}$.

Adopting this embedding strategy, the final predictions will be obtained
executing a circuit which is explicitly depending on the data:
\begin{equation}
  \bm{y}_{\rm est} = G(\bm{x}|\bm{\theta}) = \braket{ 0| \mathcal{C}^{\dagger}(\bm{x}|\bm{\theta}) \hat{O} \,
  \mathcal{C}(\bm{x}|\bm{\theta}) |0}.
\end{equation}    

Finally, there are numerous possible choices for the loss
function and the optimizer, depending on the type of model chosen and whether one 
is doing simulation or executing on a real quantum hardware. For example, by doing
simulation of a VQC, a natural choice is to select a
well known gradient-based optimizer~\cite{Rumelhart1986LearningRB, 
kingma2017adam, JMLR:v12:duchi11a, Schmidhuber_2015, ruder2017overview, robbiati2022quantum} or some 
meta-heuristic algorithm like evolutionary strategies~\cite{hansen2023cma}, simulated
annealing~\cite{inbook}, etc.
Instead, when deploying the algorithm in actual quantum hardware it can be more effective to use
shot-frugal optimizers~\cite{K_bler_2020, arrasmith2020operator, Menickelly_2023}
or to calculate gradients using metrics better suited to the QC context~\cite{Stokes_2020}.

\subsection{Circuit's ansatz}
\label{subsec:ansatz}

Building up our QML models, we encode the input data into the parametric gates 
of the circuit following the strategy suggested in~\cite{P_rez_Salinas_2020}, 
according to which an external variable can be uploaded into the angle $\phi$ of
rotational gates of the form:
\begin{equation}
  R_k(\phi) = \exp{-i \phi \hat{\sigma}_k},
  \label{eq:rotational_gate}
\end{equation}
where the hermitian generator of the rotation $\hat{\sigma}_k$ is one of the Pauli's matrices.

In particular,
we implement an architecture inspired by the uploading layer 
described in~\cite{P_rez_Salinas_2020} called  \textit{fundamental Fourier Gate}, $\mathcal{U}$,
which is composed of five sequential rotations around the $z$ and the $y$ axis.
Our $\mathcal{U}$ implementation is as follows: 
\begin{equation}
\mathcal{U}(x|\bm{\theta}) = R_z(\theta_1) R_y(\theta_2) R_z(\theta_3) 
R_z(\theta_4\, x) R_y(\theta_5),
\label{eq:Fourier_gate}
\end{equation}
where the data $x$ is uploaded into the second gate in order of application on 
the initial state. The power of this approach lies in the fact that by re-uploading
the data $x$ into $N$ consecutive channels in the form of Eq.~(\ref{eq:Fourier_gate}), 
we approximate a target continuous function as would an $N$-term Fourier series~\cite{PerezSalinas2021}. 

This strategy is introduced for a single qubit system in which a single variable 
is re-uploaded, but is easily extendible 
to a many-qubit case. Moreover, increasing the number of qubits also increases
the flexibility of the model, allowing us to upload different variables into
different wires of the circuit. Several choices of architecture can be done, and we 
present here two of the various models implemented within the code accompanying this work.

\begin{figure}
  \centering
  \includegraphics[width=1\linewidth]{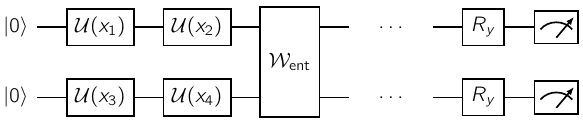}
  \caption{Diagram representation
  of the reuploading ansatz used to fit a 4-dim function. The $\mathcal{U}(x_i)$ 
  quantum channel corresponds to the fundamental 
  Fourier Gate presented in~\cite{P_rez_Salinas_2020}, while the entangling channel
  $\mathcal{W}_{\rm ent}$ is built with a combination of $CZ$ gates. This encoding 
  layer is repeated $N_{\rm layers}$ times. Finally, an $R_y$ gate is added to 
  each qubit. All parameters of the ansatz are included in the training.}
  \label{fig:reuploading}
\end{figure}

The first one is shown in Fig.~(\ref{fig:reuploading}), we encode two dimensions in every qubit,
such that the width of the circuit is equal to half the number of dimensions ($N_{\rm dim}/2$).
Each uploading of the couple of variables $(x_j, x_{j+1})$ into the associated qubit
is in the form presented in Eq.~\eqref{eq:Fourier_gate}.

Each family of gates $\{\mathcal{U}_j, \mathcal{U}_{j+1}\}$,
with $\mathcal{U}_j = \mathcal{U}(x_j)$,
is then followed by an entangling channel $\mathcal{W}_{\rm ent}$, which 
distributes the information accumulated by each qubit to the entire system. After 
the last layer, a final rotation $R_y$ is added to each wire of the circuit before 
performing the measurements.

\begin{figure}
  \centering
  \includegraphics[width=1\linewidth]{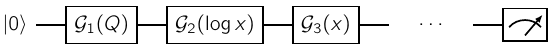}
  \caption{Schematic representation 
  of the qPDF ansatz used to fit the $u$ quark PDF. The presented gates have to
  be considered as a single layer of the ansatz, which is then repeated $N_{\rm layers}$
  times.}
  \label{fig:qpdf}
\end{figure}

Secondly, we implement a two-dimensional extension of the qPDF model from Ref.~\cite{P_rez_Salinas_2021}.
This second ansatz, is shown in Fig.~(\ref{fig:qpdf}), where we define the following channels:
\begin{equation}
\begin{cases}
\mathcal{G}_{1}(Q) = R_y(\alpha_1\, Q + \beta_1), \\ 
\mathcal{G}_{2}(\log{x}) = R_z(\alpha_2\, \log{x} + \beta_2), \\
\mathcal{G}_{3}(x) = R_y(\alpha_3\, x + \beta_3), \\
\end{cases}
\label{eq:qpdf_angles}
\end{equation}
with $\alpha_i$ and $\beta_i$ variational parameters.
The input variables $x$ and $Q$ represent, respectively, a momentum fraction and an energy, and will be more thoroughly described in the sequel.

The qPDF model is introduced in Ref.~\cite{P_rez_Salinas_2021} to fit a Parton Distribution Function (PDF)
using a VQC.
PDFs are quasi probability distributions describing the partonic (gluons and quarks) content of the proton
in hadron collisions.
For the purposes of this study, it is enough to describe them as the probability of finding a given
parton (for instance, the $u$-quark) at an energy $Q$ carrying a fraction of the total momentum of the proton $x$.
PDFs can only be computed from first principles in specific limits and thus the usual strategy is to obtain the distribution as a fit to data.
In Ref.~\cite{P_rez_Salinas_2021} the NNPDF framework~\cite{NNPDF:2021uiq} is extended from using a NN to use a VQC instead. Other approaches include Gaussian Processes~\cite{Candido:2024hjt} or polynomial forms.
For a more detailed discussion on PDFs and the techniques used in their determination, we refer the reader to Ref.~\cite{Forte:2013wc}. 

In Eq.~\eqref{eq:qpdf_angles} the imbalance of $2:1$ in uploading the $x$ and $Q$ variables into the model
follows the strategy of Ref.~\cite{NNPDF:2021njg} to capture the different behaviors of the PDF at low values of $x$ (logarithmic) and high values (linear).

This is one example of how a circuit designed to approximate a family of problems
(in this case PDFs) can also be satisfactory exploited to integrate said family of functions thanks to the parameter
shift rule, which will be briefly described in the next section.

\subsection{Derivative of a quantum circuit}
\label{subsec:psr}

Our aim is then to use the derivative of $G$ defined in Eq.~\eqref{eq:qml_predictor}
with respect to $\bm{x}$ as predictor of the integrand function presented in Eq.~\eqref{eq:integral}.
 For this we use the Parameter Shift Rule (PSR) as the method for 
calculating the derivatives of $G$ with respect to $\bm{x}$. The first example of
PSR was presented in~\cite{Mitarai_2018} and introduced a method for calculating
the derivative of an expectation value in the form of Eq.~(\ref{eq:qml_predictor})
with respect to one of the rotation angles affecting the quantum circuit $\mathcal{C}$.
We refer to a more general PSR formula presented in~\cite{Schuld_2019} and further
developed later~\cite{crooks2019gradients, Mari_2021, Banchi_2021, Wierichs_2022}.

According to~\cite{Schuld_2019}, if a circuit depends on a parameter $\mu$ (with $\mu$ a component of the vector $\bm{\theta}$) through 
a single gate $U$ whose hermitian generator has at most two eigenvalues, the derivative 
of $G$ with respect to $\mu$ can be exactly calculated as follows:
\begin{equation}
g(\mu) \equiv \partial_\mu G(\bm{\theta}) = r \bigl( G(\mu^+) - G(\mu^-) \bigr),
\label{eq:derivative_of_G}
\end{equation}
where $\pm r$ are the eigenvalues of the generator of $U$, $\mu^{\pm} = \mu \pm s$
and $s=\pi/4r$.
In this work we limit ourselves to rotational gates such as Eq.~\eqref{eq:rotational_gate} for which
$r=\frac{1}{2}$ and $s=\frac{\pi}{2}$.
The derivative of the circuit corresponds then to the execution of the same circuit
twice per gate to which the input parameter has been uploaded to.
 
Since we never upload two input parameters to the same gate,
the multidimensional extension is a trivial sequential application of the PSR per dimension and gate.
If every dimension is uploaded once to every layer ($l$),
the total number of expectation values necessary is $(2l)^{N_{\rm dim}}$.
Note that the number of input parameters does not need to coincide with the dimensionality of the integral,
making this method particularly useful for parametric integrals which are less prone to the so-called curse of dimensionality.

\subsection{Solving integrals with quantum circuits}
\label{subsec:integral_estimation}

In the following section we describe the QML training procedure and, once 
the optimization is done, how the final model can be used to calculate the 
integral of the target function.

Calculating the derivative of the circuit with the PSR, we are able to use the 
same architecture to evaluate the primitive of a function $G$ and any of its derivatives $g$.
To be more explicit, if we recall the formula with which we started the paper:
\begin{equation}
  I(\bm{\alpha}) =  \int_{\bm{x}_a}^{\bm{x}_b} g(\bm{\alpha};\bm{x})\,\dd^n\bm{x},
  \label{eq:finite_integral}
\end{equation}
the finite integral $I(\bm{\alpha})$ can also be calculated in the hypercube defined
by $(\bm{x}_{a}, \bm{x}_{b})$ by using its primitive $G(\bm{x};\bm{\alpha})$:
\begin{equation}
  I(\bm{\alpha}) = \displaystyle\sum_{x_{1},\cdots,x_{n}=x_{a}, x_{b}} (-1)^{\# a} G(x_{1},\cdots,x_{n};\bm{\alpha}), \label{eq:multidim_eval}
\end{equation}
where the sum runs over all combinations of the integration limits $a$ and $b$ and $\# a$ counts the number of variables evaluated in the lower limit $a$. In the 1-dimensional case the equation above simplifies to:
\begin{equation}
I(\bm{\alpha}) = G(x_b; \bm{\alpha}) - G(x_a; \bm{\alpha}),
\label{eq:integral_evaluation}
\end{equation}
with
\begin{equation}
G(x; \bm{\alpha}) = \int g(\bm{\alpha};x) \dd x.
\end{equation}

Our goal will be training the derivative of a VQC such that it approximates the function $g$
at any point $\bm{x}_{j}$ within the integration limits $(\bm{x}_{a}, \bm{x}_{b})$.
\begin{equation}
  g_{j, \rm est}(\bm{\alpha};\bm{x}_{j}|\bm{\theta}) = \left.
  \frac{\partial{G(\bm{\alpha}, x_1, ..., x_n|\bm{\theta})}}{\partial x_1 \,  ... \, \partial x_n}
  \right\rvert_{\bm{x}_j}.
  \label{eq:estimator_of_g}
\end{equation}

If we have a way of evaluating $g(\bm{\alpha};\bm{x})$ or have access to measurements of its value
we can generate a set of $N_{\rm train}$ training data.
Once the predictions $\{g_{j, \rm est}\}_{j=1}^{N_{\rm train}}$ are calculated for all the training data,
we quantify the goodness of our model by evaluating a Mean-Squared Error loss function:
\begin{equation}
J_{\rm mse} = \frac{1}{N_{\rm train}} 
\sum_{j=1}^{N_{\rm train}} \biggl[ g_{j, \rm meas} - 
g_{j, \rm est}(\bm{\alpha};\bm{x}_{j}|\bm{\theta})  \biggr]^2,
\label{eq:mse}
\end{equation} 
where we indicate with the index $(j)$ the $j$-th element in the training dataset 
and its associated integrand value.
The training is thus performed by iteratively updating the parameters $\bm{\theta}$ 
in order to minimize Eq.~(\ref{eq:mse}). Various optimizers have been tested, 
including Powell method~\cite{Powell1964AnEM}, L-BFGS~\cite{10.5555/3112655.3112866},
a Covariance Matrix Adaptation Evolutionary Strategy~\cite{hansen2023cma} (CMA-ES) 
and a Basin-Hopping algorithm~\cite{Wales_1997}.
Even though benchmarking different optimizers performances goes beyond the scope 
of this paper, we stress the importance of considering different optimization approaches 
when training variational quantum algorithms with and without shot-noise, namely,
simulating the sampling of the final state in the selected measurement basis. In fact, 
some of the optimizers which lead to very good results in the case of exact 
simulation (e.g. L-BFGS), become unusable when activating the shot-noise. This is 
expected, since such algorithms make use of numerical differentiation during the 
optimization process, and the numerical gradients cannot be computed in a loss function
landscape which suffers from the natural randomness provoked by the shot-noise.
To implement gradient-based optimization strategies which are compatible with 
the shot-noise computation, one should execute the calculation of the gradients using 
techniques which are more robust to the randomness of the system, such as parameter-shift
rules~\cite{Schuld_2019}. Although this can be a perfectly viable choice, we decide to 
follow different strategies in this work, not to worry about scalability problems 
like those introduced in~\cite{abbas2023quantum}.

We adopt two different optimization approaches in the two training
scenarios, using L-BFGS when optimizing in the exact simulation regime, and preferring 
heuristic algorithms like CMA-ES and Basin-Hopping when activating the shot-noise simulation.

Once the circuit's parameters are optimized, we have a fixed architecture
which is a surrogate of $G(x; \bm{\alpha})$ and
that can be used to evaluate integrals with respect to any combination of the target 
variables.
This aspect makes the strategy particularly interesting when dealing
with high-dimensional functions.

As an example, we can marginalize the integrand previously defined over the variable $x_{k}$
\begin{equation}
  I_{ab}(\dots,x_{k-1},x_{k+1},\dots) = \int_{x_{k,a}}^{x_{k,b}} g(\bm{x}) \text{d}x_k,
\label{eq:integral_marginalization}
\end{equation}
by uploading the integration limits, $x_{k, a}$ and $x_{k, b}$
and removing only that derivative from Eq.~\eqref{eq:estimator_of_g}:
\begin{equation}
\begin{split}
I_{ab}(\dots  &,x_{k-1} ,x_{k+1},\dots) \simeq \\
&  g_{\text{est}}(x_{k, b}|\bm{\theta}_{\rm best}) - g_{\text{est}}(x_{k, a}|\bm{\theta}_{\rm best}), 
\label{eq:integral_calculus}
\end{split}
\end{equation}
where we write $g_{\text{est}}$ explicitly depending only on $x_k$ for simplicity.
This can be extended for the partial integration of any of the variables of which $I$ depends on,
until finally we recover the full integration as shown in Eq.~\eqref{eq:multidim_eval}.

\section{Results}
\label{sec:results}
In order to showcase the possibilities of the methodology presented in this paper we are going to use
a VQC for two different target functions.
We will first show the flexibility of the method to obtain total or partial integrals and differential distributions,
and then we will apply to a practical case in which the approach can introduce a net-gain.
Both examples are implemented in the public code which accompanies this paper
(among other examples).

\subsection{Toy Model}
For our first example we utilize the ansatz of Ref.~\ref{fig:reuploading}
to approximate a d-dimensional trigonometric function:
\begin{equation}
g(\bm{x}) = \cos ( \bm{\alpha} \cdot \bm{x} + \alpha_{0}),
\label{eq:trigonometric_target}
\end{equation}
with $\bm{x}$ and $\bm{\alpha}$ n-dim vectors.
The integral of Eq.~\eqref{eq:trigonometric_target},
while trivial to perform analytically,
will serve to demonstrate how training one single circuit to obtain the primitive,
\begin{equation}
  I(\bm{\alpha};\bm{x}) = \displaystyle\int g(\bm{\alpha};\bm{x})  \dd \bm{x} \label{eq:integralG},
\end{equation}
can provide us the flexibility to obtain other derived quantities.
 
\begin{figure*}
  \centering
  \includegraphics[width=0.5\linewidth]{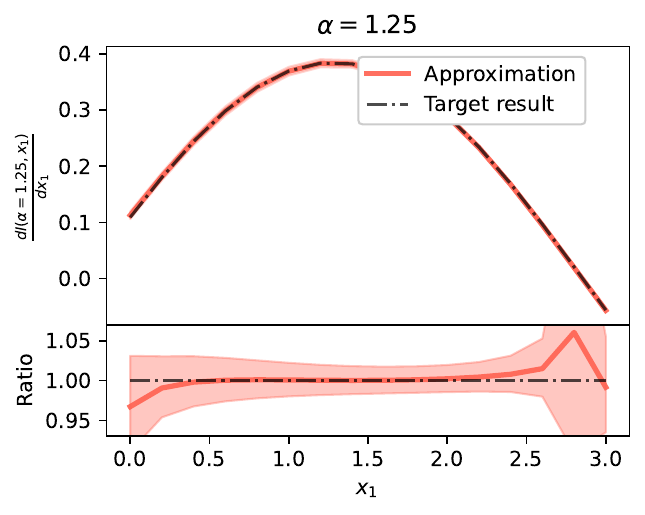}%
  \includegraphics[width=0.5\linewidth]{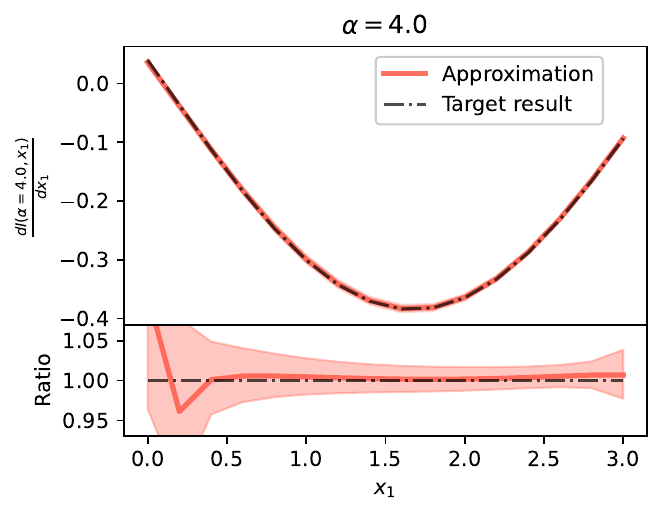}
  \caption{Differential distribution of the integral of Eq.~\eqref{eq:trigonometric_target}
  on $x_{1}$ for different values of $\alpha_0$ with $\bm{\alpha} = \{1, 2, \frac{1}{2}\}$.
  The ranges chosen for $x_{1}$ are also the ranges used for the integration of all other variables. 
  These results are obtained through exact state vector simulation.
  The training was performed with 100 points in $x$ and 10 different values of $\alpha_0 \in (0, 5)$,
  using the L-BFGS optimizer for 200-300 iterations.
  The error bands are computed retraining the circuit for different seeds.
  }
  \label{fig:cosinemarginal}
\end{figure*}

For instance, we might be interested on the differential distributions $\frac{\dd I(\bm{\alpha};\bm{x})}{\dd x_{i}}$ for a given $i$
and for different values of one of the parameters $\bm{\alpha}$.
In general, this would require to perform the numerical integration once per choice of $i$, per bin in the distribution
and choice of $\alpha$.
By having a surrogate for $I(\bm{\alpha};\bm{x})$ we can obtain each distribution as seen in Fig.~\ref{fig:cosinemarginal}, 
were we collect results obtained by training the model with exact state vector 
simulation of the quantum circuits.

In Fig.~\ref{fig:cosinemarginal} we have plotted the differential distribution
\begin{equation}
  \frac{\dd I(\bm{\alpha};x_{1})}{\dd x_{1}},
\end{equation}
for two different values of $\alpha_0$ within the training range $(0, 5)$.
All other parameters have remained fixed in order to minimize the computing cost in this toy-model example.
The training range for the integrated variables $(x_{1}, x_{2}, x_{3})$ has been $(0, 3.5)$.
In the plots we choose to integrate $x_{2}$ and $x_{3}$ from 0 to 3 for every value fo $x_{1}$,
but any choice of integration limits within the integration range would be possible.

A shortcoming of this approach is the lack of an uncertainty associated to the numerical integration.
In Ref.~\cite{Ma_tre_2023} the suggestion is to use an ensemble of replicas of the network trained to the same
data in order to use the variance as an error.
We have followed the same strategy here by training an ensemble of circuits randomizing the choice of training
points which leads to a spread of the results.

Other numerical methods provide some shortcuts to obtain similar results.
For instance MC integration methods would allow us to bin quantities which depend on the integration variables
(provided that we know beforehand the distributions that we want to obtain).
However, a change in the parameter $\alpha_0$ will always lead to a new integration.
Instead, once we have a circuit that approximates Eq.~\eqref{eq:integralG}, any derived quantity in the training range is accessible
without any new runs.

It is important to remark that there is no free lunch,
we are paying the penalty in terms of evaluations of the integrand (and the surrogate) during the training,
but the outcome is a flexible representation of the final quantity which can then be reutilized.
Similarly to Monte Carlo methods,
the accuracy of the calculation can be improved by increasing computational cost with
either a larger number of samples or longer training lengths.
Note that in this case we have a function that enables us to generate an unlimited amount of data for arbitrary inputs and so the limitation
is only of computational cost.

\subsection{The $u$-quark PDF}
\label{sec:uquark}

In the previous section we have chosen an ideal scenario with a function for which we know the primitive and against which we can exactly test.
In what follows we consider an actual use-case for the approach that we propose in this paper:
the integration of a function which is only known numerically
and for which we have a representation in the form of a VQC.

For this we use the PDF fit and corresponding ansatz presented in Ref.~\cite{P_rez_Salinas_2021}.
Note that in order to overcome some of the computational challenges that arose when swapping the
NN for a VQC, in Ref.~\cite{P_rez_Salinas_2021} the resulting PDF is not normalized.
Normalizing the PDF requires taking the integral over $x$ at the fixed fitting energy ($Q_{0}$),
which in turn requires many evaluations of the PDF at every stage of the fit.

Instead, with the techniques introduced in this paper we use the ansatz described in Sec.~\ref{subsec:ansatz} to obtain a model by which we can produce both the $u$-quark PDF
and its own integral, hence allowing for a prediction normalized by construction without the need for an expensive numerical integral.

\begin{figure}
  \centering
  \includegraphics[width=1\linewidth]{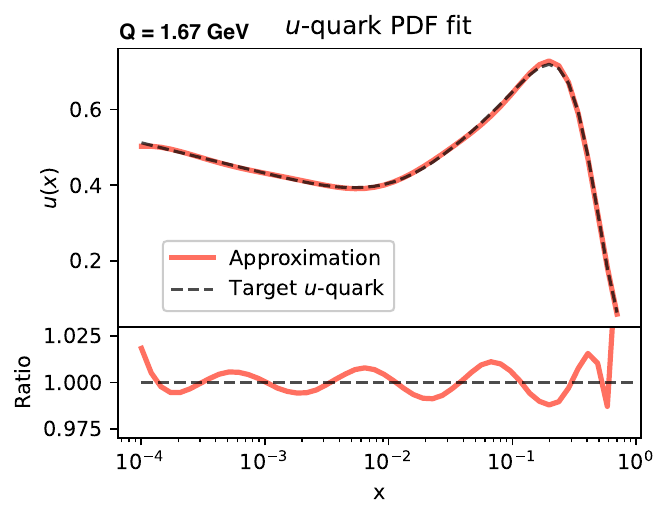}
  \includegraphics[width=1\linewidth]{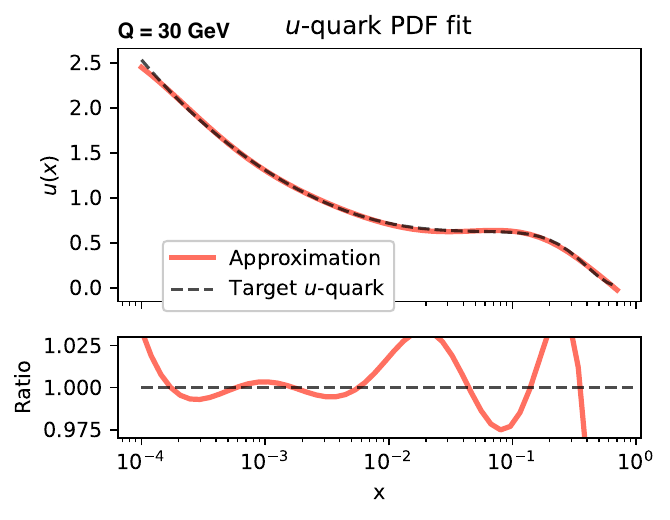}
  \caption{\label{fig:uquark_fit_fixed_q} Comparison between the fit of the derivative of the circuit (red line) and the target result read directly from the interpolation grids for the central NNPDF4.0 replica for the $u$-quark for a fixed value of $Q=1.67$ GeV (fitting scale of NNDPF4.0)
  and $Q = 30$ GeV.
  While the ansatz in Sec.~\ref{subsec:ansatz} gives us a model for the integral, the results training is performed
  onto the derivative.
  The training set uses approx. 100 points for each of the 5 fixed values of $Q$ chosen between the initial $Q=1.67$ GeV and $Q=40$ GeV.
  These results are obtained through exact state vector simulation.
  In Fig.~\ref{fig:uquark_fit} we will use the same strategy,
  (training over a wider range of $Q$) to plot the integrand as a function of the energy.}
\end{figure}

In Fig.~\ref{fig:uquark_fit_fixed_q} we show the result of fitting the derivative of the ansatz to the training data for a fixed value of $Q$,
obtaining a very good description across the entire range.

We train the model with the L-BFGS optimizer and performing exact simulation using \texttt{Qibo}.
As the training data we utilize directly the $u$-quark PDF from the NNPDF4.0~\cite{NNPDF:2021njg} PDF grids.
We limit the training to the $x$ range (1e-4, 0.7) which is comparable with the ranges of data available in PDF determinations.
In Fig.~\ref{fig:uquark_fit_fixed_q} we train for several fixed values of $Q$ chosen linearly between $1.65$ and $40$ GeV. 

Once we are able to approximate the integrand with the derivative of the ansatz, the associated integral,
\begin{equation}
  I_{u}(Q) = \int_{10^{-4}}^{0.7} x\, u(x,Q) \, \text{d}x, 
\label{eq:uquark2d_integral}
\end{equation}
is the ingredient necessary for the normalization of the $u$-quark PDF.

Note that the $Q$-dependence of the PDF can be computed analytically given the PDF at a fixed scale $Q_0$,
thus in a fit only a value of $I_{u}(Q_0)$ is required.
The ability to train the VQC for a range of values of $Q$ can be exploited to evaluate Eq.~\eqref{eq:uquark2d_integral}
as a parametric integral for a range of values of $Q$.

In order to show the generalizability potential of the method we have also trained the circuit for a wide range of values of $Q$,
chosen such that no quark-threshold is crossed (defined as the value of the energy for which the number 
of active quarks in NNPDF4.0 changes) to reduce instabilities.

In QML, in real-life scenarios,
one needs to account for the probabilistic shot-noise associated with the measurement of the quantum states
and the noise associated to the hardware.
While the uncertainty associated to the training shown in Fig.~\ref{fig:cosinemarginal}
can in principle be reduced by increasing the training length,
these uncertainties are intrinsic to the methodology.

\begin{figure}
  \centering
  \includegraphics[width=1\linewidth]{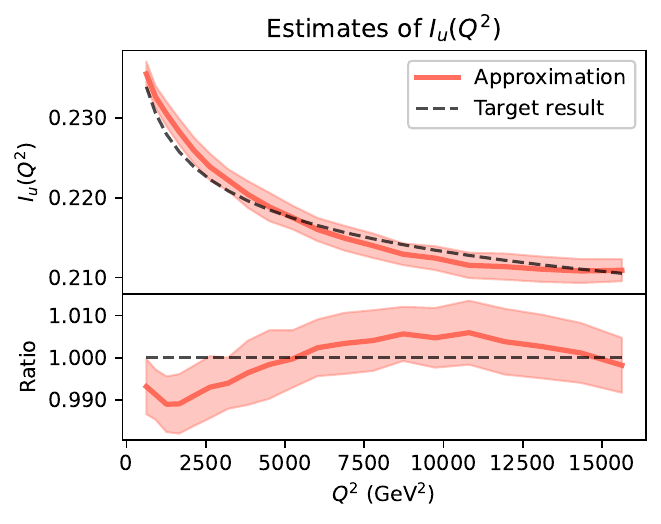}
  \caption{\label{fig:uquark_fit} Above, integral of $x\,u(x,Q)$ calculated for $N_q=20$ values of $Q$.
  These results are obtained by simulating circuits with shot-noise. In particular, 
  for each $q$ we perform $N_{\rm runs}=100$ predictions and each prediction is 
  obtained by executing the circuit $N_{\rm shots}=10^{6}$ times. The orange line 
  and its confidence belt are calculated using mean and one standard deviation 
  over the $N_{\rm runs}$ prediction sets. Below, average relative percentage 
  error calculated using the $N_{\rm runs}$ predictions.
  The training has been performed on approx. 100 different values of $Q$
  evenly spaced on $Q^{2}$ and took about 20 h in a 32-cores machine.
  }
\end{figure}

In Fig.~\ref{fig:uquark_fit} we show the calculation of the integral of Eq.~\eqref{eq:uquark2d_integral}
for different values of $Q$ within the training range.
The circuits are simulated with shot-noise since we perform $N_{\rm shots}=10^6$ 
to compute each expected value (circuit is called $N_{\rm shots}$ times for each 
estimation of the primitive $G$). We then repeat every measurement of the integration
$N_{\rm runs}=100$ times.
The error is computed by taking the standard deviation $\sigma$ of the measurement.
The shaded band in Fig.~\ref{fig:uquark_fit} correspond to the 1$\sigma$ band.

This corresponds to an ideal real-life scenario since we are not considering hardware noise.
The shot-noise error scales as $\frac{1}{\sqrt{N_{\tt runs}}}$.
This statistical noise leads in this case to
an uncertainty of about $1\%$.
Due to the computational cost we don't include in this case the training uncertainty,
which would need to be included (possibly added in quadrature).
We also note a systematic shift with respect to the true value of $\sim$ 5\textperthousand).

\subsection{Normalized by construction}

In this section we detail an application of our method to an actual physics problem
for which the improvement with respect to classical approaches (e.g. numerical integration using Monte Carlo methods)
is immediate: the determination of PDFs using quantum computers~\cite{P_rez_Salinas_2021}.
We leave the actual implementation to future work, but we outline here the methodology
and expected gains.
Note that the same strategy can be applied whenever the integral of the function is part of the
fitting process.

In Refs.~\cite{NNPDF:2021njg,NNPDF:2021uiq} the PDFs are parametrized at $Q_{0}=1.67$ GeV such that 
\begin{equation}
    V(x) = \frac{3\hat{V}(x)}{\displaystyle\int_{x_a}^{x_b} \dd x \hat{V}(x)},\label{eq:sumrules}
\end{equation}
where $V(x)$ ($\hat{V}$) corresponds to the valence (unnormalized) PDF in the so-called ``evolution basis''.
and which can be written in terms of the quark and antiquark PDFs as,
\begin{equation}
  V(x) = \displaystyle\sum^{u,d,s,c} q(x) -\bar{q}(x).
\end{equation}
Note that at the parametrization energy only the lighter quarks ($u$, $d$, $s$ and $c$) PDFs are independent
while the heavier $b$ and $t$ quarks can be determined from the quarks (and gluon) PDFs.

The integral in the denominator of Eq.~\eqref{eq:sumrules} is computed numerically over $x$ between
the extremes $x_{a}$ and $x_{b}$ and is computed from scratch for every training step.
Consequently, obtaining one single value for $V(x)$ requires a costly numerical integral
(more than $10^3$ function calls) for each step of the training process.

With the \texttt{QiNNtegrate} approach we can instead construct a PDF which is
already normalized:
\begin{equation}
  V(x) =\frac{\hat{V}}{I_{\hat{V}}} = \frac{3\displaystyle\sum^{\text{shifts}}_{s}G(x_{s})}{G(x_b) -G(x_a)},
\end{equation}
where we have exchanged the computational burden of computing the numerical integrand by
a costlier evaluation of the unnormalized distribution.
The shifts in the sum corresponds to all pairs of $(x_{+}, x_{-})$ needed to compute
the derivative as per Eq.~\eqref{eq:derivative_of_G}.

For the ansatz used in this work we would need $16$ circuit executions per point in $x$ involved in the fit
to estimate the derivative but have removed a number of executions in the order of $10^3$ from the training process.
In the concrete case of Ref.~\cite{NNPDF:2021njg} this can result in a net reduction of function calls of a factor of approximately 6.
Furthermore, the fit would in this case benefit from a simpler functional form.

In addition, novel proposals for non-demolition measurements can be integrated in 
this algorithm to further reduce the number of shifts required to estimate the 
integrand function. To give an example, in~\cite{Solinas:2023nvb, minuto2024novel} 
the authors introduce an algorithm to reduce the number of expectation values required
to reconstruct the gradient formula originally introduced in~\cite{Schuld_2019}. 
This is done by exploiting ancillary qubits and encoding the required shifts into a single 
circuit. 

Together with the improvement of the qubit's quality, we believe such techniques 
can help in making our proposed integration algorithm even more useful in practice.

\subsection{Integrating on a real qubit}
\label{sec:hardware}
In this section we present some results obtained executing this algorithm on a 
real quantum hardware. In particular, we use a superconducting device composed 
of a single qubit hosted at the Quantum Research Center (QRC) of the 
Technology Innovation Institute (TII). The entire process is realized using the 
\Qibo~\cite{Efthymiou_2021, Efthymiou_2022,
Carrazza_2023, pasquale2023opensource, stavros_efthymiou_2023_7736837, 
stavros_efthymiou_2023_7748527, andrea_pasquale_2023_7662185}
 ecosystem; the high level code is written with \Qibo
and then executed on the qubit by \Qibolab~\cite{stavros_efthymiou_2023_7748527}. 
In case we make use of \Qibosoq~\cite{rodolfo_carobene_2023_8126172}, which is 
the server that integrates Qick~\cite{stefanazzi2022qick} in the \Qibolab ecosystem for executing arbitrary 
circuits and pulse sequences through RFsoC FPGA boards. All the single qubit characterization and calibration 
routines are performed using \Qibocal~\cite{andrea_pasquale_2023_7662185,pasquale2023opensource}.  
\begin{figure}
  \centering
  \includegraphics[width=1\linewidth]{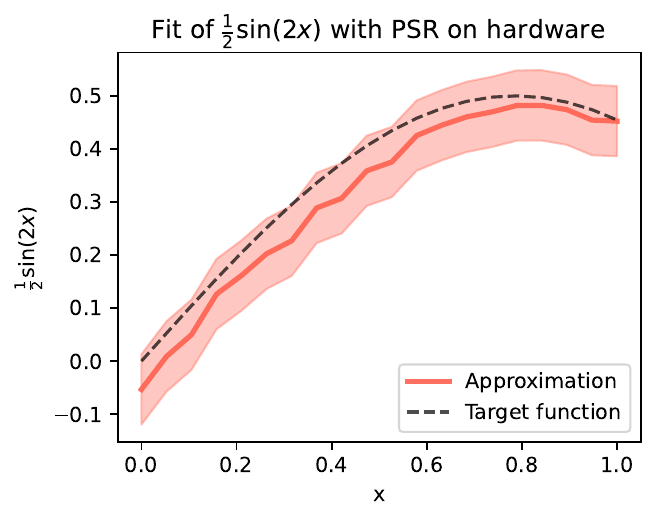}
  \caption{Estimates of the integrand $g(x)=\frac{1}{2}\sin(2x)$ between $x=0$ and $x=1$ obtained
  by executing the presented algorithm on a real superconducting qubit. The target 
  $N_{\rm data}=20$ function values (black dashed line) are compared with the estimates 
  (orange line), obtained as the average of $N_{\rm runs}=5$ sets of predictions. The confidence
  interval is drawn using $2\sigma$ error over the experiments. The results are computed
  without any kind of error mitigation technique.}
  \label{fig:hardware}
\end{figure}

We tackle a simple example to reduce the number of expected values of the form of 
Eq.~\eqref{eq:qml_predictor} to be evaluated. In fact, since the derivative of 
a circuit is used as predictor in our model, $2*N_{x}$ expected values
are needed to compute each estimation, where $N_x$ is the number of times $x$ 
is uploaded into the model. For example, in case of the one dimensional $u$ quark PDF presented
in Sec.~\ref{sec:uquark} $N_x=2\,N_{\rm layers}$. We present in Fig.~\ref{fig:hardware} 
predictions of $N_{\rm data}=20$ values of the integrand $g(x)=\frac{1}{2}\sin(2x)$, considering 
$N_{\rm data}$ values of $x$ uniformly distributed in $[0,1]$. We calculate $N_{\rm runs}=5$ times 
the prediction for each $x$ and use the mean and $2\sigma$ to define the confidence 
intervals around the estimates which are shown in Fig.~\ref{fig:hardware}. 
During the process, each expectation value
is obtained executing the circuit $N_{\rm shots}=5000$ times.
This simple example proves a simple integrand function can be fitted on a NISQ
device. 
 
As second test, we calculate the integral value of the target function:
\begin{equation}
I_{\rm target} = \int_0^1 \frac{1}{2}\sin{(2x)}\dd x
\end{equation}
$N_{\rm int}=10$ times obtaining as estimate: $\hat{I}_{\rm target} = 0.326 \pm 0.011$,
to be compared with the exact value $I_{\rm target} = 0.354$.
The error on the estimate is the standard deviation over the $N_{\rm int}$ results.
We believe having such a satisfactory result even with noisy hardware
can be motivated by the nature of the problem we are 
tackling. In fact, the target integral is here calculated following Eq.~\eqref{eq:integral_evaluation},
and the difference between estimations can help in removing systematic errors 
that may occur when dealing with NISQ devices.

\section{Conclusion}
\label{sec:conclusion}

In this paper we have extended the methods proposed in
Refs.~\cite{DBLP:journals/corr/abs-2012-01714, Ma_tre_2023}
to quantum computers and shown how the properties of these new type of devices can introduce a practical advantage
compared to classical alternatives by exploiting the properties of the parameter shift rule.

Furthermore, we have demonstrated a practical-case in which one can obtain a net benefit by utilizing this approach,
by skipping entirely the need for a numerical integration during a fitting procedure.
A natural extension of this work is an enhancement of the methodology proposed in Ref.~\cite{P_rez_Salinas_2021}
in which the PDF fit could not be normalized due to technical constraints.

We then reported interesting results obtained on a real superconducting qubit, showing 
how a NISQ device can already be used to fit integrand functions following our 
algorithm. 
We have also made all code available in a public python framework \texttt{QiNNtegrate}~\cite{QiNNtegrate}.
which can be used to reproduce the results of this work and which can be extended
to other custom functions.
\texttt{QiNNtegrate} is based on \texttt{Qibo} and thus the generated circuits can be either simulated
in a classical computer or directly executed on hardware.

\acknowledgments We thank D. Maitre for the careful reading of the manuscript and many very useful comments.
This project is supported by CERN's Quantum Technology
Initiative (QTI). MR is supported by CERN doctoral program. SC thanks the TH
hospitality during the elaboration of this manuscript.

\bibliography{references}

\begin{thebibliography}{73}
\expandafter\ifx\csname natexlab\endcsname\relax\def\natexlab#1{#1}\fi
\expandafter\ifx\csname bibnamefont\endcsname\relax
  \def\bibnamefont#1{#1}\fi
\expandafter\ifx\csname bibfnamefont\endcsname\relax
  \def\bibfnamefont#1{#1}\fi
\expandafter\ifx\csname citenamefont\endcsname\relax
  \def\citenamefont#1{#1}\fi
\expandafter\ifx\csname url\endcsname\relax
  \def\url#1{\texttt{#1}}\fi
\expandafter\ifx\csname urlprefix\endcsname\relax\def\urlprefix{URL }\fi
\providecommand{\bibinfo}[2]{#2}
\providecommand{\eprint}[2][]{\url{#2}}

\bibitem[{\citenamefont{Metropolis and Ulam}(1949)}]{10.2307/2280232}
\bibinfo{author}{\bibfnamefont{N.}~\bibnamefont{Metropolis}} \bibnamefont{and}
  \bibinfo{author}{\bibfnamefont{S.}~\bibnamefont{Ulam}},
  \bibinfo{journal}{Journal of the American Statistical Association}
  \textbf{\bibinfo{volume}{44}}, \bibinfo{pages}{335} (\bibinfo{year}{1949}),
  ISSN \bibinfo{issn}{01621459},
  \urlprefix\url{http://www.jstor.org/stable/2280232}.

\bibitem[{\citenamefont{Caflisch}(1998)}]{caflisch_1998}
\bibinfo{author}{\bibfnamefont{R.~E.} \bibnamefont{Caflisch}},
  \bibinfo{journal}{Acta Numerica} \textbf{\bibinfo{volume}{7}},
  \bibinfo{pages}{1–49} (\bibinfo{year}{1998}).

\bibitem[{\citenamefont{Zhong and Feng}(2022)}]{zhong2022efficient}
\bibinfo{author}{\bibfnamefont{H.}~\bibnamefont{Zhong}} \bibnamefont{and}
  \bibinfo{author}{\bibfnamefont{X.}~\bibnamefont{Feng}},
  \emph{\bibinfo{title}{An efficient and fast sparse grid algorithm for
  high-dimensional numerical integration}} (\bibinfo{year}{2022}),
  \eprint{2210.14313}.

\bibitem[{\citenamefont{Ghahramani and Rasmussen}(2002)}]{NIPS2002_24917db1}
\bibinfo{author}{\bibfnamefont{Z.}~\bibnamefont{Ghahramani}} \bibnamefont{and}
  \bibinfo{author}{\bibfnamefont{C.}~\bibnamefont{Rasmussen}}, in
  \emph{\bibinfo{booktitle}{Advances in Neural Information Processing
  Systems}}, edited by
  \bibinfo{editor}{\bibfnamefont{S.}~\bibnamefont{Becker}},
  \bibinfo{editor}{\bibfnamefont{S.}~\bibnamefont{Thrun}}, \bibnamefont{and}
  \bibinfo{editor}{\bibfnamefont{K.}~\bibnamefont{Obermayer}}
  (\bibinfo{publisher}{MIT Press}, \bibinfo{year}{2002}),
  vol.~\bibinfo{volume}{15},
  \urlprefix\url{https://proceedings.neurips.cc/paper_files/paper/2002/file/24917db15c4e37e421866448c9ab23d8-Paper.pdf}.

\bibitem[{\citenamefont{Schmidhuber}(2015{\natexlab{a}})}]{SCHMIDHUBER201585}
\bibinfo{author}{\bibfnamefont{J.}~\bibnamefont{Schmidhuber}},
  \bibinfo{journal}{Neural Networks} \textbf{\bibinfo{volume}{61}},
  \bibinfo{pages}{85} (\bibinfo{year}{2015}{\natexlab{a}}), ISSN
  \bibinfo{issn}{0893-6080},
  \urlprefix\url{https://www.sciencedirect.com/science/article/pii/S0893608014002135}.

\bibitem[{\citenamefont{Lepage}(1978)}]{Lepage:1977sw}
\bibinfo{author}{\bibfnamefont{G.~P.} \bibnamefont{Lepage}},
  \bibinfo{journal}{J. Comput. Phys.} \textbf{\bibinfo{volume}{27}},
  \bibinfo{pages}{192} (\bibinfo{year}{1978}).

\bibitem[{\citenamefont{Lepage}(2021)}]{Lepage:2020tgj}
\bibinfo{author}{\bibfnamefont{G.~P.} \bibnamefont{Lepage}},
  \bibinfo{journal}{J. Comput. Phys.} \textbf{\bibinfo{volume}{439}},
  \bibinfo{pages}{110386} (\bibinfo{year}{2021}), \eprint{2009.05112}.

\bibitem[{\citenamefont{Carrazza and Cruz-Martinez}(2020)}]{Carrazza:2020rdn}
\bibinfo{author}{\bibfnamefont{S.}~\bibnamefont{Carrazza}} \bibnamefont{and}
  \bibinfo{author}{\bibfnamefont{J.~M.} \bibnamefont{Cruz-Martinez}},
  \bibinfo{journal}{Comput. Phys. Commun.} \textbf{\bibinfo{volume}{254}},
  \bibinfo{pages}{107376} (\bibinfo{year}{2020}), \eprint{2002.12921}.

\bibitem[{\citenamefont{G\'omez et~al.}(2021)\citenamefont{G\'omez, Toftevaag,
  and Meoni}}]{Gomez:2021czl}
\bibinfo{author}{\bibfnamefont{P.}~\bibnamefont{G\'omez}},
  \bibinfo{author}{\bibfnamefont{H.~H.} \bibnamefont{Toftevaag}},
  \bibnamefont{and} \bibinfo{author}{\bibfnamefont{G.}~\bibnamefont{Meoni}},
  \bibinfo{journal}{J. Open Source Softw.} \textbf{\bibinfo{volume}{6}},
  \bibinfo{pages}{3439} (\bibinfo{year}{2021}).

\bibitem[{\citenamefont{Kleiss and Pittau}(1994)}]{Kleiss:1994qy}
\bibinfo{author}{\bibfnamefont{R.}~\bibnamefont{Kleiss}} \bibnamefont{and}
  \bibinfo{author}{\bibfnamefont{R.}~\bibnamefont{Pittau}},
  \bibinfo{journal}{Comput. Phys. Commun.} \textbf{\bibinfo{volume}{83}},
  \bibinfo{pages}{141} (\bibinfo{year}{1994}), \eprint{hep-ph/9405257}.

\bibitem[{\citenamefont{M\"{u}ller et~al.}(2019)\citenamefont{M\"{u}ller,
  Mcwilliams, Rousselle, Gross, and Nov\'{a}k}}]{10.1145/3341156}
\bibinfo{author}{\bibfnamefont{T.}~\bibnamefont{M\"{u}ller}},
  \bibinfo{author}{\bibfnamefont{B.}~\bibnamefont{Mcwilliams}},
  \bibinfo{author}{\bibfnamefont{F.}~\bibnamefont{Rousselle}},
  \bibinfo{author}{\bibfnamefont{M.}~\bibnamefont{Gross}}, \bibnamefont{and}
  \bibinfo{author}{\bibfnamefont{J.}~\bibnamefont{Nov\'{a}k}},
  \bibinfo{journal}{ACM Trans. Graph.} \textbf{\bibinfo{volume}{38}}
  (\bibinfo{year}{2019}), ISSN \bibinfo{issn}{0730-0301},
  \urlprefix\url{https://doi.org/10.1145/3341156}.

\bibitem[{\citenamefont{Bothmann et~al.}(2020)\citenamefont{Bothmann,
  Jan\ss{}en, Knobbe, Schmale, and Schumann}}]{Bothmann:2020ywa}
\bibinfo{author}{\bibfnamefont{E.}~\bibnamefont{Bothmann}},
  \bibinfo{author}{\bibfnamefont{T.}~\bibnamefont{Jan\ss{}en}},
  \bibinfo{author}{\bibfnamefont{M.}~\bibnamefont{Knobbe}},
  \bibinfo{author}{\bibfnamefont{T.}~\bibnamefont{Schmale}}, \bibnamefont{and}
  \bibinfo{author}{\bibfnamefont{S.}~\bibnamefont{Schumann}},
  \bibinfo{journal}{SciPost Phys.} \textbf{\bibinfo{volume}{8}},
  \bibinfo{pages}{069} (\bibinfo{year}{2020}), \eprint{2001.05478}.

\bibitem[{\citenamefont{Lindell et~al.}(2020)\citenamefont{Lindell, Martel, and
  Wetzstein}}]{DBLP:journals/corr/abs-2012-01714}
\bibinfo{author}{\bibfnamefont{D.~B.} \bibnamefont{Lindell}},
  \bibinfo{author}{\bibfnamefont{J.~N.~P.} \bibnamefont{Martel}},
  \bibnamefont{and}
  \bibinfo{author}{\bibfnamefont{G.}~\bibnamefont{Wetzstein}},
  \bibinfo{journal}{CoRR} \textbf{\bibinfo{volume}{abs/2012.01714}}
  (\bibinfo{year}{2020}), \eprint{2012.01714},
  \urlprefix\url{https://arxiv.org/abs/2012.01714}.

\bibitem[{\citenamefont{Ma{\^{\i}}tre and Santos-Mateos}(2023)}]{Ma_tre_2023}
\bibinfo{author}{\bibfnamefont{D.}~\bibnamefont{Ma{\^{\i}}tre}}
  \bibnamefont{and}
  \bibinfo{author}{\bibfnamefont{R.}~\bibnamefont{Santos-Mateos}},
  \bibinfo{journal}{Journal of High Energy Physics}
  \textbf{\bibinfo{volume}{2023}} (\bibinfo{year}{2023}),
  \urlprefix\url{https://doi.org/10.1007%2Fjhep03%282023%29221}.

\bibitem[{\citenamefont{Schuld et~al.}(2014)\citenamefont{Schuld, Sinayskiy,
  and Petruccione}}]{Schuld_2014}
\bibinfo{author}{\bibfnamefont{M.}~\bibnamefont{Schuld}},
  \bibinfo{author}{\bibfnamefont{I.}~\bibnamefont{Sinayskiy}},
  \bibnamefont{and}
  \bibinfo{author}{\bibfnamefont{F.}~\bibnamefont{Petruccione}},
  \bibinfo{journal}{Contemporary Physics} \textbf{\bibinfo{volume}{56}},
  \bibinfo{pages}{172} (\bibinfo{year}{2014}),
  \urlprefix\url{https://doi.org/10.1080%2F00107514.2014.964942}.

\bibitem[{\citenamefont{Biamonte et~al.}(2017)\citenamefont{Biamonte, Wittek,
  Pancotti, Rebentrost, Wiebe, and Lloyd}}]{Biamonte_2017}
\bibinfo{author}{\bibfnamefont{J.}~\bibnamefont{Biamonte}},
  \bibinfo{author}{\bibfnamefont{P.}~\bibnamefont{Wittek}},
  \bibinfo{author}{\bibfnamefont{N.}~\bibnamefont{Pancotti}},
  \bibinfo{author}{\bibfnamefont{P.}~\bibnamefont{Rebentrost}},
  \bibinfo{author}{\bibfnamefont{N.}~\bibnamefont{Wiebe}}, \bibnamefont{and}
  \bibinfo{author}{\bibfnamefont{S.}~\bibnamefont{Lloyd}},
  \bibinfo{journal}{Nature} \textbf{\bibinfo{volume}{549}},
  \bibinfo{pages}{195} (\bibinfo{year}{2017}),
  \urlprefix\url{https://doi.org/10.1038%2Fnature23474}.

\bibitem[{\citenamefont{Mitarai et~al.}(2018)\citenamefont{Mitarai, Negoro,
  Kitagawa, and Fujii}}]{Mitarai_2018}
\bibinfo{author}{\bibfnamefont{K.}~\bibnamefont{Mitarai}},
  \bibinfo{author}{\bibfnamefont{M.}~\bibnamefont{Negoro}},
  \bibinfo{author}{\bibfnamefont{M.}~\bibnamefont{Kitagawa}}, \bibnamefont{and}
  \bibinfo{author}{\bibfnamefont{K.}~\bibnamefont{Fujii}},
  \bibinfo{journal}{Physical Review A} \textbf{\bibinfo{volume}{98}}
  (\bibinfo{year}{2018}),
  \urlprefix\url{https://doi.org/10.1103%2Fphysreva.98.032309}.

\bibitem[{\citenamefont{Chen et~al.}(2020)\citenamefont{Chen, Yang, Qi, Chen,
  Ma, and Goan}}]{chen2020variational}
\bibinfo{author}{\bibfnamefont{S.~Y.-C.} \bibnamefont{Chen}},
  \bibinfo{author}{\bibfnamefont{C.-H.~H.} \bibnamefont{Yang}},
  \bibinfo{author}{\bibfnamefont{J.}~\bibnamefont{Qi}},
  \bibinfo{author}{\bibfnamefont{P.-Y.} \bibnamefont{Chen}},
  \bibinfo{author}{\bibfnamefont{X.}~\bibnamefont{Ma}}, \bibnamefont{and}
  \bibinfo{author}{\bibfnamefont{H.-S.} \bibnamefont{Goan}},
  \emph{\bibinfo{title}{Variational quantum circuits for deep reinforcement
  learning}} (\bibinfo{year}{2020}), \eprint{1907.00397}.

\bibitem[{\citenamefont{Abbas et~al.}(2021)\citenamefont{Abbas, Sutter, Zoufal,
  Lucchi, Figalli, and Woerner}}]{Abbas_2021}
\bibinfo{author}{\bibfnamefont{A.}~\bibnamefont{Abbas}},
  \bibinfo{author}{\bibfnamefont{D.}~\bibnamefont{Sutter}},
  \bibinfo{author}{\bibfnamefont{C.}~\bibnamefont{Zoufal}},
  \bibinfo{author}{\bibfnamefont{A.}~\bibnamefont{Lucchi}},
  \bibinfo{author}{\bibfnamefont{A.}~\bibnamefont{Figalli}}, \bibnamefont{and}
  \bibinfo{author}{\bibfnamefont{S.}~\bibnamefont{Woerner}},
  \bibinfo{journal}{Nature Computational Science} \textbf{\bibinfo{volume}{1}},
  \bibinfo{pages}{403} (\bibinfo{year}{2021}),
  \urlprefix\url{https://doi.org/10.1038%2Fs43588-021-00084-1}.

\bibitem[{\citenamefont{Schuld et~al.}(2019)\citenamefont{Schuld, Bergholm,
  Gogolin, Izaac, and Killoran}}]{Schuld_2019}
\bibinfo{author}{\bibfnamefont{M.}~\bibnamefont{Schuld}},
  \bibinfo{author}{\bibfnamefont{V.}~\bibnamefont{Bergholm}},
  \bibinfo{author}{\bibfnamefont{C.}~\bibnamefont{Gogolin}},
  \bibinfo{author}{\bibfnamefont{J.}~\bibnamefont{Izaac}}, \bibnamefont{and}
  \bibinfo{author}{\bibfnamefont{N.}~\bibnamefont{Killoran}},
  \bibinfo{journal}{Physical Review A} \textbf{\bibinfo{volume}{99}}
  (\bibinfo{year}{2019}),
  \urlprefix\url{https://doi.org/10.1103%2Fphysreva.99.032331}.

\bibitem[{\citenamefont{Crooks}(2019)}]{crooks2019gradients}
\bibinfo{author}{\bibfnamefont{G.~E.} \bibnamefont{Crooks}},
  \emph{\bibinfo{title}{Gradients of parameterized quantum gates using the
  parameter-shift rule and gate decomposition}} (\bibinfo{year}{2019}),
  \eprint{1905.13311}.

\bibitem[{\citenamefont{Mari et~al.}(2021)\citenamefont{Mari, Bromley, and
  Killoran}}]{Mari_2021}
\bibinfo{author}{\bibfnamefont{A.}~\bibnamefont{Mari}},
  \bibinfo{author}{\bibfnamefont{T.~R.} \bibnamefont{Bromley}},
  \bibnamefont{and} \bibinfo{author}{\bibfnamefont{N.}~\bibnamefont{Killoran}},
  \bibinfo{journal}{Physical Review A} \textbf{\bibinfo{volume}{103}}
  (\bibinfo{year}{2021}),
  \urlprefix\url{https://doi.org/10.1103%2Fphysreva.103.012405}.

\bibitem[{\citenamefont{Wierichs et~al.}(2022)\citenamefont{Wierichs, Izaac,
  Wang, and Lin}}]{Wierichs_2022}
\bibinfo{author}{\bibfnamefont{D.}~\bibnamefont{Wierichs}},
  \bibinfo{author}{\bibfnamefont{J.}~\bibnamefont{Izaac}},
  \bibinfo{author}{\bibfnamefont{C.}~\bibnamefont{Wang}}, \bibnamefont{and}
  \bibinfo{author}{\bibfnamefont{C.~Y.-Y.} \bibnamefont{Lin}},
  \bibinfo{journal}{Quantum} \textbf{\bibinfo{volume}{6}}, \bibinfo{pages}{677}
  (\bibinfo{year}{2022}),
  \urlprefix\url{https://doi.org/10.22331%2Fq-2022-03-30-677}.

\bibitem[{\citenamefont{Efthymiou et~al.}(2021)\citenamefont{Efthymiou,
  Ramos-Calderer, Bravo-Prieto, P{\'{e}}rez-Salinas,
  Garc{\'{\i}}a-Mart{\'{\i}}n, Garcia-Saez, Latorre, and
  Carrazza}}]{Efthymiou_2021}
\bibinfo{author}{\bibfnamefont{S.}~\bibnamefont{Efthymiou}},
  \bibinfo{author}{\bibfnamefont{S.}~\bibnamefont{Ramos-Calderer}},
  \bibinfo{author}{\bibfnamefont{C.}~\bibnamefont{Bravo-Prieto}},
  \bibinfo{author}{\bibfnamefont{A.}~\bibnamefont{P{\'{e}}rez-Salinas}},
  \bibinfo{author}{\bibfnamefont{D.}~\bibnamefont{Garc{\'{\i}}a-Mart{\'{\i}}n}},
  \bibinfo{author}{\bibfnamefont{A.}~\bibnamefont{Garcia-Saez}},
  \bibinfo{author}{\bibfnamefont{J.~I.} \bibnamefont{Latorre}},
  \bibnamefont{and} \bibinfo{author}{\bibfnamefont{S.}~\bibnamefont{Carrazza}},
  \bibinfo{journal}{Quantum Science and Technology}
  \textbf{\bibinfo{volume}{7}}, \bibinfo{pages}{015018} (\bibinfo{year}{2021}),
  \urlprefix\url{https://doi.org/10.1088%2F2058-9565%2Fac39f5}.

\bibitem[{\citenamefont{Efthymiou et~al.}(2022)\citenamefont{Efthymiou,
  Lazzarin, Pasquale, and Carrazza}}]{Efthymiou_2022}
\bibinfo{author}{\bibfnamefont{S.}~\bibnamefont{Efthymiou}},
  \bibinfo{author}{\bibfnamefont{M.}~\bibnamefont{Lazzarin}},
  \bibinfo{author}{\bibfnamefont{A.}~\bibnamefont{Pasquale}}, \bibnamefont{and}
  \bibinfo{author}{\bibfnamefont{S.}~\bibnamefont{Carrazza}},
  \bibinfo{journal}{Quantum} \textbf{\bibinfo{volume}{6}}, \bibinfo{pages}{814}
  (\bibinfo{year}{2022}),
  \urlprefix\url{https://doi.org/10.22331%2Fq-2022-09-22-814}.

\bibitem[{\citenamefont{Carrazza et~al.}(2023)\citenamefont{Carrazza,
  Efthymiou, Lazzarin, and Pasquale}}]{Carrazza_2023}
\bibinfo{author}{\bibfnamefont{S.}~\bibnamefont{Carrazza}},
  \bibinfo{author}{\bibfnamefont{S.}~\bibnamefont{Efthymiou}},
  \bibinfo{author}{\bibfnamefont{M.}~\bibnamefont{Lazzarin}}, \bibnamefont{and}
  \bibinfo{author}{\bibfnamefont{A.}~\bibnamefont{Pasquale}},
  \bibinfo{journal}{Journal of Physics: Conference Series}
  \textbf{\bibinfo{volume}{2438}}, \bibinfo{pages}{012148}
  (\bibinfo{year}{2023}),
  \urlprefix\url{https://doi.org/10.1088%2F1742-6596%2F2438%2F1%2F012148}.

\bibitem[{\citenamefont{Pasquale
  et~al.}(2023{\natexlab{a}})\citenamefont{Pasquale, Efthymiou, Ramos-Calderer,
  Wilkens, Roth, and Carrazza}}]{pasquale2023opensource}
\bibinfo{author}{\bibfnamefont{A.}~\bibnamefont{Pasquale}},
  \bibinfo{author}{\bibfnamefont{S.}~\bibnamefont{Efthymiou}},
  \bibinfo{author}{\bibfnamefont{S.}~\bibnamefont{Ramos-Calderer}},
  \bibinfo{author}{\bibfnamefont{J.}~\bibnamefont{Wilkens}},
  \bibinfo{author}{\bibfnamefont{I.}~\bibnamefont{Roth}}, \bibnamefont{and}
  \bibinfo{author}{\bibfnamefont{S.}~\bibnamefont{Carrazza}},
  \emph{\bibinfo{title}{Towards an open-source framework to perform quantum
  calibration and characterization}} (\bibinfo{year}{2023}{\natexlab{a}}),
  \eprint{2303.10397}.

\bibitem[{\citenamefont{Efthymiou
  et~al.}(2023{\natexlab{a}})}]{stavros_efthymiou_2023_7736837}
\bibinfo{author}{\bibfnamefont{S.}~\bibnamefont{Efthymiou}}
  \bibnamefont{et~al.}, \emph{\bibinfo{title}{qiboteam/qibo: Qibo 0.1.12}}
  (\bibinfo{year}{2023}{\natexlab{a}}),
  \urlprefix\url{https://doi.org/10.5281/zenodo.7736837}.

\bibitem[{\citenamefont{Efthymiou
  et~al.}(2023{\natexlab{b}})}]{stavros_efthymiou_2023_7748527}
\bibinfo{author}{\bibfnamefont{S.}~\bibnamefont{Efthymiou}}
  \bibnamefont{et~al.}, \emph{\bibinfo{title}{qiboteam/qibolab: Qibolab 0.0.2}}
  (\bibinfo{year}{2023}{\natexlab{b}}),
  \urlprefix\url{https://doi.org/10.5281/zenodo.7748527}.

\bibitem[{\citenamefont{Pasquale
  et~al.}(2023{\natexlab{b}})}]{andrea_pasquale_2023_7662185}
\bibinfo{author}{\bibfnamefont{A.}~\bibnamefont{Pasquale}}
  \bibnamefont{et~al.}, \emph{\bibinfo{title}{qiboteam/qibocal: Qibocal 0.0.1}}
  (\bibinfo{year}{2023}{\natexlab{b}}),
  \urlprefix\url{https://doi.org/10.5281/zenodo.7662185}.

\bibitem[{\citenamefont{Preskill}(2018)}]{Preskill_2018}
\bibinfo{author}{\bibfnamefont{J.}~\bibnamefont{Preskill}},
  \bibinfo{journal}{Quantum} \textbf{\bibinfo{volume}{2}}, \bibinfo{pages}{79}
  (\bibinfo{year}{2018}),
  \urlprefix\url{https://doi.org/10.22331%2Fq-2018-08-06-79}.

\bibitem[{\citenamefont{Delgado et~al.}(2022)\citenamefont{Delgado, Hamilton,
  Date, Vlimant, Magano, Omar, Bargassa, Francis, Gianelle, Sestini
  et~al.}}]{delgado2022quantum}
\bibinfo{author}{\bibfnamefont{A.}~\bibnamefont{Delgado}},
  \bibinfo{author}{\bibfnamefont{K.~E.} \bibnamefont{Hamilton}},
  \bibinfo{author}{\bibfnamefont{P.}~\bibnamefont{Date}},
  \bibinfo{author}{\bibfnamefont{J.-R.} \bibnamefont{Vlimant}},
  \bibinfo{author}{\bibfnamefont{D.}~\bibnamefont{Magano}},
  \bibinfo{author}{\bibfnamefont{Y.}~\bibnamefont{Omar}},
  \bibinfo{author}{\bibfnamefont{P.}~\bibnamefont{Bargassa}},
  \bibinfo{author}{\bibfnamefont{A.}~\bibnamefont{Francis}},
  \bibinfo{author}{\bibfnamefont{A.}~\bibnamefont{Gianelle}},
  \bibinfo{author}{\bibfnamefont{L.}~\bibnamefont{Sestini}},
  \bibnamefont{et~al.}, \emph{\bibinfo{title}{Quantum computing for data
  analysis in high energy physics}} (\bibinfo{year}{2022}),
  \eprint{2203.08805}.

\bibitem[{\citenamefont{Gustafson et~al.}(2022)\citenamefont{Gustafson,
  Prestel, Spannowsky, and Williams}}]{Gustafson:2022dsq}
\bibinfo{author}{\bibfnamefont{G.}~\bibnamefont{Gustafson}},
  \bibinfo{author}{\bibfnamefont{S.}~\bibnamefont{Prestel}},
  \bibinfo{author}{\bibfnamefont{M.}~\bibnamefont{Spannowsky}},
  \bibnamefont{and} \bibinfo{author}{\bibfnamefont{S.}~\bibnamefont{Williams}},
  \bibinfo{journal}{JHEP} \textbf{\bibinfo{volume}{11}}, \bibinfo{pages}{035}
  (\bibinfo{year}{2022}), \eprint{2207.10694}.

\bibitem[{\citenamefont{Agliardi et~al.}(2022)\citenamefont{Agliardi, Grossi,
  Pellen, and Prati}}]{Agliardi:2022ghn}
\bibinfo{author}{\bibfnamefont{G.}~\bibnamefont{Agliardi}},
  \bibinfo{author}{\bibfnamefont{M.}~\bibnamefont{Grossi}},
  \bibinfo{author}{\bibfnamefont{M.}~\bibnamefont{Pellen}}, \bibnamefont{and}
  \bibinfo{author}{\bibfnamefont{E.}~\bibnamefont{Prati}},
  \bibinfo{journal}{Phys. Lett. B} \textbf{\bibinfo{volume}{832}},
  \bibinfo{pages}{137228} (\bibinfo{year}{2022}), \eprint{2201.01547}.

\bibitem[{\citenamefont{Bauer et~al.}(2023)}]{Bauer:2022hpo}
\bibinfo{author}{\bibfnamefont{C.~W.} \bibnamefont{Bauer}}
  \bibnamefont{et~al.}, \bibinfo{journal}{PRX Quantum}
  \textbf{\bibinfo{volume}{4}}, \bibinfo{pages}{027001} (\bibinfo{year}{2023}),
  \eprint{2204.03381}.

\bibitem[{\citenamefont{Woźniak et~al.}(2023)\citenamefont{Woźniak, Belis,
  Puljak, Barkoutsos, Dissertori, Grossi, Pierini, Reiter, Tavernelli, and
  Vallecorsa}}]{wozniak2023quantum}
\bibinfo{author}{\bibfnamefont{K.~A.} \bibnamefont{Woźniak}},
  \bibinfo{author}{\bibfnamefont{V.}~\bibnamefont{Belis}},
  \bibinfo{author}{\bibfnamefont{E.}~\bibnamefont{Puljak}},
  \bibinfo{author}{\bibfnamefont{P.}~\bibnamefont{Barkoutsos}},
  \bibinfo{author}{\bibfnamefont{G.}~\bibnamefont{Dissertori}},
  \bibinfo{author}{\bibfnamefont{M.}~\bibnamefont{Grossi}},
  \bibinfo{author}{\bibfnamefont{M.}~\bibnamefont{Pierini}},
  \bibinfo{author}{\bibfnamefont{F.}~\bibnamefont{Reiter}},
  \bibinfo{author}{\bibfnamefont{I.}~\bibnamefont{Tavernelli}},
  \bibnamefont{and}
  \bibinfo{author}{\bibfnamefont{S.}~\bibnamefont{Vallecorsa}},
  \emph{\bibinfo{title}{Quantum anomaly detection in the latent space of proton
  collision events at the lhc}} (\bibinfo{year}{2023}), \eprint{2301.10780}.

\bibitem[{\citenamefont{Chawdhry and Pellen}(2023)}]{Chawdhry:2023jks}
\bibinfo{author}{\bibfnamefont{H.~A.} \bibnamefont{Chawdhry}} \bibnamefont{and}
  \bibinfo{author}{\bibfnamefont{M.}~\bibnamefont{Pellen}}
  (\bibinfo{year}{2023}), \eprint{2303.04818}.

\bibitem[{\citenamefont{Robbiati et~al.}(2023)\citenamefont{Robbiati,
  Cruz-Martinez, and Carrazza}}]{robbiati2023determining}
\bibinfo{author}{\bibfnamefont{M.}~\bibnamefont{Robbiati}},
  \bibinfo{author}{\bibfnamefont{J.~M.} \bibnamefont{Cruz-Martinez}},
  \bibnamefont{and} \bibinfo{author}{\bibfnamefont{S.}~\bibnamefont{Carrazza}},
  \emph{\bibinfo{title}{Determining probability density functions with
  adiabatic quantum computing}} (\bibinfo{year}{2023}), \eprint{2303.11346}.

\bibitem[{\citenamefont{D'Elia et~al.}(2024)}]{DElia:2024pax}
\bibinfo{author}{\bibfnamefont{A.}~\bibnamefont{D'Elia}} \bibnamefont{et~al.},
  \bibinfo{journal}{Appl. Sciences} \textbf{\bibinfo{volume}{14}},
  \bibinfo{pages}{1478} (\bibinfo{year}{2024}), \eprint{2402.04322}.

\bibitem[{\citenamefont{Benedetti et~al.}(2019)\citenamefont{Benedetti, Lloyd,
  Sack, and Fiorentini}}]{Benedetti_2019}
\bibinfo{author}{\bibfnamefont{M.}~\bibnamefont{Benedetti}},
  \bibinfo{author}{\bibfnamefont{E.}~\bibnamefont{Lloyd}},
  \bibinfo{author}{\bibfnamefont{S.}~\bibnamefont{Sack}}, \bibnamefont{and}
  \bibinfo{author}{\bibfnamefont{M.}~\bibnamefont{Fiorentini}},
  \bibinfo{journal}{Quantum Science and Technology}
  \textbf{\bibinfo{volume}{4}}, \bibinfo{pages}{043001} (\bibinfo{year}{2019}),
  \urlprefix\url{https://doi.org/10.1088%2F2058-9565%2Fab4eb5}.

\bibitem[{\citenamefont{Cerezo et~al.}(2020)\citenamefont{Cerezo, Arrasmith,
  Babbush, Benjamin, Endo, Fujii, McClean, Mitarai, Yuan, Cincio
  et~al.}}]{cerezo2020variational}
\bibinfo{author}{\bibfnamefont{M.}~\bibnamefont{Cerezo}},
  \bibinfo{author}{\bibfnamefont{A.}~\bibnamefont{Arrasmith}},
  \bibinfo{author}{\bibfnamefont{R.}~\bibnamefont{Babbush}},
  \bibinfo{author}{\bibfnamefont{S.~C.} \bibnamefont{Benjamin}},
  \bibinfo{author}{\bibfnamefont{S.}~\bibnamefont{Endo}},
  \bibinfo{author}{\bibfnamefont{K.}~\bibnamefont{Fujii}},
  \bibinfo{author}{\bibfnamefont{J.~R.} \bibnamefont{McClean}},
  \bibinfo{author}{\bibfnamefont{K.}~\bibnamefont{Mitarai}},
  \bibinfo{author}{\bibfnamefont{X.}~\bibnamefont{Yuan}},
  \bibinfo{author}{\bibfnamefont{L.}~\bibnamefont{Cincio}},
  \bibnamefont{et~al.}, \emph{\bibinfo{title}{Variational quantum algorithms}}
  (\bibinfo{year}{2020}), \bibinfo{note}{cite arxiv:2012.09265Comment: Review
  Article. 29 pages, 6 figures},
  \urlprefix\url{http://arxiv.org/abs/2012.09265}.

\bibitem[{\citenamefont{Lloyd et~al.}(2020)\citenamefont{Lloyd, Schuld, Ijaz,
  Izaac, and Killoran}}]{lloyd2020quantum}
\bibinfo{author}{\bibfnamefont{S.}~\bibnamefont{Lloyd}},
  \bibinfo{author}{\bibfnamefont{M.}~\bibnamefont{Schuld}},
  \bibinfo{author}{\bibfnamefont{A.}~\bibnamefont{Ijaz}},
  \bibinfo{author}{\bibfnamefont{J.}~\bibnamefont{Izaac}}, \bibnamefont{and}
  \bibinfo{author}{\bibfnamefont{N.}~\bibnamefont{Killoran}},
  \emph{\bibinfo{title}{Quantum embeddings for machine learning}}
  (\bibinfo{year}{2020}), \eprint{2001.03622}.

\bibitem[{\citenamefont{P{\'{e}}rez-Salinas
  et~al.}(2020)\citenamefont{P{\'{e}}rez-Salinas, Cervera-Lierta, Gil-Fuster,
  and Latorre}}]{P_rez_Salinas_2020}
\bibinfo{author}{\bibfnamefont{A.}~\bibnamefont{P{\'{e}}rez-Salinas}},
  \bibinfo{author}{\bibfnamefont{A.}~\bibnamefont{Cervera-Lierta}},
  \bibinfo{author}{\bibfnamefont{E.}~\bibnamefont{Gil-Fuster}},
  \bibnamefont{and} \bibinfo{author}{\bibfnamefont{J.~I.}
  \bibnamefont{Latorre}}, \bibinfo{journal}{Quantum}
  \textbf{\bibinfo{volume}{4}}, \bibinfo{pages}{226} (\bibinfo{year}{2020}),
  \urlprefix\url{https://doi.org/10.22331%2Fq-2020-02-06-226}.

\bibitem[{\citenamefont{Incudini et~al.}(2022)\citenamefont{Incudini, Martini,
  and Pierro}}]{incudini2022structure}
\bibinfo{author}{\bibfnamefont{M.}~\bibnamefont{Incudini}},
  \bibinfo{author}{\bibfnamefont{F.}~\bibnamefont{Martini}}, \bibnamefont{and}
  \bibinfo{author}{\bibfnamefont{A.~D.} \bibnamefont{Pierro}},
  \emph{\bibinfo{title}{Structure learning of quantum embeddings}}
  (\bibinfo{year}{2022}), \eprint{2209.11144}.

\bibitem[{\citenamefont{Schuld and Petruccione}(2018)}]{Schuld:2018gao}
\bibinfo{author}{\bibfnamefont{M.}~\bibnamefont{Schuld}} \bibnamefont{and}
  \bibinfo{author}{\bibfnamefont{F.}~\bibnamefont{Petruccione}},
  \emph{\bibinfo{title}{{Supervised Learning with Quantum Computers}}}, Quantum
  Science and Technology (\bibinfo{publisher}{Springer}, \bibinfo{year}{2018}).

\bibitem[{\citenamefont{Rumelhart et~al.}(1986)\citenamefont{Rumelhart, Hinton,
  and Williams}}]{Rumelhart1986LearningRB}
\bibinfo{author}{\bibfnamefont{D.~E.} \bibnamefont{Rumelhart}},
  \bibinfo{author}{\bibfnamefont{G.~E.} \bibnamefont{Hinton}},
  \bibnamefont{and} \bibinfo{author}{\bibfnamefont{R.~J.}
  \bibnamefont{Williams}}, \bibinfo{journal}{Nature}
  \textbf{\bibinfo{volume}{323}}, \bibinfo{pages}{533} (\bibinfo{year}{1986}).

\bibitem[{\citenamefont{Kingma and Ba}(2017)}]{kingma2017adam}
\bibinfo{author}{\bibfnamefont{D.~P.} \bibnamefont{Kingma}} \bibnamefont{and}
  \bibinfo{author}{\bibfnamefont{J.}~\bibnamefont{Ba}},
  \emph{\bibinfo{title}{Adam: A method for stochastic optimization}}
  (\bibinfo{year}{2017}), \eprint{1412.6980}.

\bibitem[{\citenamefont{Duchi et~al.}(2011)\citenamefont{Duchi, Hazan, and
  Singer}}]{JMLR:v12:duchi11a}
\bibinfo{author}{\bibfnamefont{J.}~\bibnamefont{Duchi}},
  \bibinfo{author}{\bibfnamefont{E.}~\bibnamefont{Hazan}}, \bibnamefont{and}
  \bibinfo{author}{\bibfnamefont{Y.}~\bibnamefont{Singer}},
  \bibinfo{journal}{Journal of Machine Learning Research}
  \textbf{\bibinfo{volume}{12}}, \bibinfo{pages}{2121} (\bibinfo{year}{2011}),
  \urlprefix\url{http://jmlr.org/papers/v12/duchi11a.html}.

\bibitem[{\citenamefont{Schmidhuber}(2015{\natexlab{b}})}]{Schmidhuber_2015}
\bibinfo{author}{\bibfnamefont{J.}~\bibnamefont{Schmidhuber}},
  \bibinfo{journal}{Neural Networks} \textbf{\bibinfo{volume}{61}},
  \bibinfo{pages}{85} (\bibinfo{year}{2015}{\natexlab{b}}),
  \urlprefix\url{https://doi.org/10.1016%2Fj.neunet.2014.09.003}.

\bibitem[{\citenamefont{Ruder}(2017)}]{ruder2017overview}
\bibinfo{author}{\bibfnamefont{S.}~\bibnamefont{Ruder}},
  \emph{\bibinfo{title}{An overview of gradient descent optimization
  algorithms}} (\bibinfo{year}{2017}), \eprint{1609.04747}.

\bibitem[{\citenamefont{Robbiati et~al.}(2022)\citenamefont{Robbiati,
  Efthymiou, Pasquale, and Carrazza}}]{robbiati2022quantum}
\bibinfo{author}{\bibfnamefont{M.}~\bibnamefont{Robbiati}},
  \bibinfo{author}{\bibfnamefont{S.}~\bibnamefont{Efthymiou}},
  \bibinfo{author}{\bibfnamefont{A.}~\bibnamefont{Pasquale}}, \bibnamefont{and}
  \bibinfo{author}{\bibfnamefont{S.}~\bibnamefont{Carrazza}},
  \emph{\bibinfo{title}{A quantum analytical adam descent through parameter
  shift rule using qibo}} (\bibinfo{year}{2022}), \eprint{2210.10787}.

\bibitem[{\citenamefont{Hansen}(2023)}]{hansen2023cma}
\bibinfo{author}{\bibfnamefont{N.}~\bibnamefont{Hansen}},
  \emph{\bibinfo{title}{The cma evolution strategy: A tutorial}}
  (\bibinfo{year}{2023}), \eprint{1604.00772}.

\bibitem[{\citenamefont{Henderson et~al.}(2006)\citenamefont{Henderson,
  Jacobson, and Johnson}}]{inbook}
\bibinfo{author}{\bibfnamefont{D.}~\bibnamefont{Henderson}},
  \bibinfo{author}{\bibfnamefont{S.}~\bibnamefont{Jacobson}}, \bibnamefont{and}
  \bibinfo{author}{\bibfnamefont{A.}~\bibnamefont{Johnson}},
  \emph{\bibinfo{title}{The Theory and Practice of Simulated Annealing}}
  (\bibinfo{year}{2006}), pp. \bibinfo{pages}{287--319}.

\bibitem[{\citenamefont{Kübler et~al.}(2020)\citenamefont{Kübler, Arrasmith,
  Cincio, and Coles}}]{K_bler_2020}
\bibinfo{author}{\bibfnamefont{J.~M.} \bibnamefont{Kübler}},
  \bibinfo{author}{\bibfnamefont{A.}~\bibnamefont{Arrasmith}},
  \bibinfo{author}{\bibfnamefont{L.}~\bibnamefont{Cincio}}, \bibnamefont{and}
  \bibinfo{author}{\bibfnamefont{P.~J.} \bibnamefont{Coles}},
  \bibinfo{journal}{Quantum} \textbf{\bibinfo{volume}{4}}, \bibinfo{pages}{263}
  (\bibinfo{year}{2020}),
  \urlprefix\url{https://doi.org/10.22331%2Fq-2020-05-11-263}.

\bibitem[{\citenamefont{Arrasmith et~al.}(2020)\citenamefont{Arrasmith, Cincio,
  Somma, and Coles}}]{arrasmith2020operator}
\bibinfo{author}{\bibfnamefont{A.}~\bibnamefont{Arrasmith}},
  \bibinfo{author}{\bibfnamefont{L.}~\bibnamefont{Cincio}},
  \bibinfo{author}{\bibfnamefont{R.~D.} \bibnamefont{Somma}}, \bibnamefont{and}
  \bibinfo{author}{\bibfnamefont{P.~J.} \bibnamefont{Coles}},
  \emph{\bibinfo{title}{Operator sampling for shot-frugal optimization in
  variational algorithms}} (\bibinfo{year}{2020}), \eprint{2004.06252}.

\bibitem[{\citenamefont{Menickelly et~al.}(2023)\citenamefont{Menickelly, Ha,
  and Otten}}]{Menickelly_2023}
\bibinfo{author}{\bibfnamefont{M.}~\bibnamefont{Menickelly}},
  \bibinfo{author}{\bibfnamefont{Y.}~\bibnamefont{Ha}}, \bibnamefont{and}
  \bibinfo{author}{\bibfnamefont{M.}~\bibnamefont{Otten}},
  \bibinfo{journal}{Quantum} \textbf{\bibinfo{volume}{7}}, \bibinfo{pages}{949}
  (\bibinfo{year}{2023}),
  \urlprefix\url{https://doi.org/10.22331%2Fq-2023-03-16-949}.

\bibitem[{\citenamefont{Stokes et~al.}(2020)\citenamefont{Stokes, Izaac,
  Killoran, and Carleo}}]{Stokes_2020}
\bibinfo{author}{\bibfnamefont{J.}~\bibnamefont{Stokes}},
  \bibinfo{author}{\bibfnamefont{J.}~\bibnamefont{Izaac}},
  \bibinfo{author}{\bibfnamefont{N.}~\bibnamefont{Killoran}}, \bibnamefont{and}
  \bibinfo{author}{\bibfnamefont{G.}~\bibnamefont{Carleo}},
  \bibinfo{journal}{Quantum} \textbf{\bibinfo{volume}{4}}, \bibinfo{pages}{269}
  (\bibinfo{year}{2020}),
  \urlprefix\url{https://doi.org/10.22331%2Fq-2020-05-25-269}.

\bibitem[{\citenamefont{Pérez-Salinas
  et~al.}(2021)\citenamefont{Pérez-Salinas, López-Núñez, García-Sáez,
  Forn-Díaz, and Latorre}}]{PerezSalinas2021}
\bibinfo{author}{\bibfnamefont{A.}~\bibnamefont{Pérez-Salinas}},
  \bibinfo{author}{\bibfnamefont{D.}~\bibnamefont{López-Núñez}},
  \bibinfo{author}{\bibfnamefont{A.}~\bibnamefont{García-Sáez}},
  \bibinfo{author}{\bibfnamefont{P.}~\bibnamefont{Forn-Díaz}},
  \bibnamefont{and} \bibinfo{author}{\bibfnamefont{J.~I.}
  \bibnamefont{Latorre}}, \bibinfo{journal}{Physical Review A}
  \textbf{\bibinfo{volume}{104}} (\bibinfo{year}{2021}), ISSN
  \bibinfo{issn}{2469-9934},
  \urlprefix\url{http://dx.doi.org/10.1103/PhysRevA.104.012405}.

\bibitem[{\citenamefont{P{\'{e}}rez-Salinas
  et~al.}(2021)\citenamefont{P{\'{e}}rez-Salinas, Cruz-Martinez, Alhajri, and
  Carrazza}}]{P_rez_Salinas_2021}
\bibinfo{author}{\bibfnamefont{A.}~\bibnamefont{P{\'{e}}rez-Salinas}},
  \bibinfo{author}{\bibfnamefont{J.}~\bibnamefont{Cruz-Martinez}},
  \bibinfo{author}{\bibfnamefont{A.~A.} \bibnamefont{Alhajri}},
  \bibnamefont{and} \bibinfo{author}{\bibfnamefont{S.}~\bibnamefont{Carrazza}},
  \bibinfo{journal}{Physical Review D} \textbf{\bibinfo{volume}{103}}
  (\bibinfo{year}{2021}),
  \urlprefix\url{https://doi.org/10.1103%2Fphysrevd.103.034027}.

\bibitem[{\citenamefont{Ball et~al.}(2021)}]{NNPDF:2021uiq}
\bibinfo{author}{\bibfnamefont{R.~D.} \bibnamefont{Ball}} \bibnamefont{et~al.}
  (\bibinfo{collaboration}{NNPDF}), \bibinfo{journal}{Eur. Phys. J. C}
  \textbf{\bibinfo{volume}{81}}, \bibinfo{pages}{958} (\bibinfo{year}{2021}),
  \eprint{2109.02671}.

\bibitem[{\citenamefont{Candido et~al.}(2024)\citenamefont{Candido, Del~Debbio,
  Giani, and Petrillo}}]{Candido:2024hjt}
\bibinfo{author}{\bibfnamefont{A.}~\bibnamefont{Candido}},
  \bibinfo{author}{\bibfnamefont{L.}~\bibnamefont{Del~Debbio}},
  \bibinfo{author}{\bibfnamefont{T.}~\bibnamefont{Giani}}, \bibnamefont{and}
  \bibinfo{author}{\bibfnamefont{G.}~\bibnamefont{Petrillo}}
  (\bibinfo{year}{2024}), \eprint{2404.07573}.

\bibitem[{\citenamefont{Forte and Watt}(2013)}]{Forte:2013wc}
\bibinfo{author}{\bibfnamefont{S.}~\bibnamefont{Forte}} \bibnamefont{and}
  \bibinfo{author}{\bibfnamefont{G.}~\bibnamefont{Watt}},
  \bibinfo{journal}{Ann. Rev. Nucl. Part. Sci.} \textbf{\bibinfo{volume}{63}},
  \bibinfo{pages}{291} (\bibinfo{year}{2013}), \eprint{1301.6754}.

\bibitem[{\citenamefont{Ball et~al.}(2022)}]{NNPDF:2021njg}
\bibinfo{author}{\bibfnamefont{R.~D.} \bibnamefont{Ball}} \bibnamefont{et~al.}
  (\bibinfo{collaboration}{NNPDF}), \bibinfo{journal}{Eur. Phys. J. C}
  \textbf{\bibinfo{volume}{82}}, \bibinfo{pages}{428} (\bibinfo{year}{2022}),
  \eprint{2109.02653}.

\bibitem[{\citenamefont{Banchi and Crooks}(2021)}]{Banchi_2021}
\bibinfo{author}{\bibfnamefont{L.}~\bibnamefont{Banchi}} \bibnamefont{and}
  \bibinfo{author}{\bibfnamefont{G.~E.} \bibnamefont{Crooks}},
  \bibinfo{journal}{Quantum} \textbf{\bibinfo{volume}{5}}, \bibinfo{pages}{386}
  (\bibinfo{year}{2021}),
  \urlprefix\url{https://doi.org/10.22331%2Fq-2021-01-25-386}.

\bibitem[{\citenamefont{Powell}(1964)}]{Powell1964AnEM}
\bibinfo{author}{\bibfnamefont{M.~J.~D.} \bibnamefont{Powell}},
  \bibinfo{journal}{Comput. J.} \textbf{\bibinfo{volume}{7}},
  \bibinfo{pages}{155} (\bibinfo{year}{1964}).

\bibitem[{\citenamefont{Liu and Nocedal}(1989)}]{10.5555/3112655.3112866}
\bibinfo{author}{\bibfnamefont{D.~C.} \bibnamefont{Liu}} \bibnamefont{and}
  \bibinfo{author}{\bibfnamefont{J.}~\bibnamefont{Nocedal}},
  \bibinfo{journal}{Math. Program.} \textbf{\bibinfo{volume}{45}}
  (\bibinfo{year}{1989}), ISSN \bibinfo{issn}{0025-5610}.

\bibitem[{\citenamefont{Wales and Doye}(1997)}]{Wales_1997}
\bibinfo{author}{\bibfnamefont{D.~J.} \bibnamefont{Wales}} \bibnamefont{and}
  \bibinfo{author}{\bibfnamefont{J.~P.~K.} \bibnamefont{Doye}},
  \bibinfo{journal}{The Journal of Physical Chemistry A}
  \textbf{\bibinfo{volume}{101}}, \bibinfo{pages}{5111} (\bibinfo{year}{1997}),
  \urlprefix\url{https://doi.org/10.1021%2Fjp970984n}.

\bibitem[{\citenamefont{Abbas et~al.}(2023)\citenamefont{Abbas, King, Huang,
  Huggins, Movassagh, Gilboa, and McClean}}]{abbas2023quantum}
\bibinfo{author}{\bibfnamefont{A.}~\bibnamefont{Abbas}},
  \bibinfo{author}{\bibfnamefont{R.}~\bibnamefont{King}},
  \bibinfo{author}{\bibfnamefont{H.-Y.} \bibnamefont{Huang}},
  \bibinfo{author}{\bibfnamefont{W.~J.} \bibnamefont{Huggins}},
  \bibinfo{author}{\bibfnamefont{R.}~\bibnamefont{Movassagh}},
  \bibinfo{author}{\bibfnamefont{D.}~\bibnamefont{Gilboa}}, \bibnamefont{and}
  \bibinfo{author}{\bibfnamefont{J.~R.} \bibnamefont{McClean}},
  \emph{\bibinfo{title}{On quantum backpropagation, information reuse, and
  cheating measurement collapse}} (\bibinfo{year}{2023}), \eprint{2305.13362}.

\bibitem[{\citenamefont{Solinas et~al.}(2023)\citenamefont{Solinas, Caletti,
  and Minuto}}]{Solinas:2023nvb}
\bibinfo{author}{\bibfnamefont{P.}~\bibnamefont{Solinas}},
  \bibinfo{author}{\bibfnamefont{S.}~\bibnamefont{Caletti}}, \bibnamefont{and}
  \bibinfo{author}{\bibfnamefont{G.}~\bibnamefont{Minuto}},
  \bibinfo{journal}{Eur. Phys. J. D} \textbf{\bibinfo{volume}{77}},
  \bibinfo{pages}{76} (\bibinfo{year}{2023}), \eprint{2301.07128}.

\bibitem[{\citenamefont{Minuto et~al.}(2024)\citenamefont{Minuto, Caletti, and
  Solinas}}]{minuto2024novel}
\bibinfo{author}{\bibfnamefont{G.}~\bibnamefont{Minuto}},
  \bibinfo{author}{\bibfnamefont{S.}~\bibnamefont{Caletti}}, \bibnamefont{and}
  \bibinfo{author}{\bibfnamefont{P.}~\bibnamefont{Solinas}},
  \emph{\bibinfo{title}{A novel approach to reduce derivative costs in
  variational quantum algorithms}} (\bibinfo{year}{2024}), \eprint{2404.02245}.

\bibitem[{\citenamefont{Carobene et~al.}(2023)\citenamefont{Carobene, Candido,
  Serrano, Carrazza, and Edoardo-Pedicillo}}]{rodolfo_carobene_2023_8126172}
\bibinfo{author}{\bibfnamefont{R.}~\bibnamefont{Carobene}},
  \bibinfo{author}{\bibfnamefont{A.}~\bibnamefont{Candido}},
  \bibinfo{author}{\bibfnamefont{J.}~\bibnamefont{Serrano}},
  \bibinfo{author}{\bibfnamefont{S.}~\bibnamefont{Carrazza}}, \bibnamefont{and}
  \bibinfo{author}{\bibnamefont{Edoardo-Pedicillo}},
  \emph{\bibinfo{title}{qiboteam/qibosoq: Qibosoq 0.0.3}}
  (\bibinfo{year}{2023}),
  \urlprefix\url{https://doi.org/10.5281/zenodo.8126172}.

\bibitem[{\citenamefont{Stefanazzi et~al.}(2022)\citenamefont{Stefanazzi,
  Treptow, Wilcer, Stoughton, Montella, Bradford, Cancelo, Saxena, Arnaldi,
  Sussman et~al.}}]{stefanazzi2022qick}
\bibinfo{author}{\bibfnamefont{L.}~\bibnamefont{Stefanazzi}},
  \bibinfo{author}{\bibfnamefont{K.}~\bibnamefont{Treptow}},
  \bibinfo{author}{\bibfnamefont{N.}~\bibnamefont{Wilcer}},
  \bibinfo{author}{\bibfnamefont{C.}~\bibnamefont{Stoughton}},
  \bibinfo{author}{\bibfnamefont{S.}~\bibnamefont{Montella}},
  \bibinfo{author}{\bibfnamefont{C.}~\bibnamefont{Bradford}},
  \bibinfo{author}{\bibfnamefont{G.}~\bibnamefont{Cancelo}},
  \bibinfo{author}{\bibfnamefont{S.}~\bibnamefont{Saxena}},
  \bibinfo{author}{\bibfnamefont{H.}~\bibnamefont{Arnaldi}},
  \bibinfo{author}{\bibfnamefont{S.}~\bibnamefont{Sussman}},
  \bibnamefont{et~al.}, \emph{\bibinfo{title}{The qick (quantum instrumentation
  control kit): Readout and control for qubits and detectors}}
  (\bibinfo{year}{2022}), \eprint{2110.00557}.

\bibitem[{\citenamefont{Martinez et~al.}(2023)\citenamefont{Martinez, Robbiati,
  and Carrazza}}]{QiNNtegrate}
\bibinfo{author}{\bibfnamefont{J.~M.~C.} \bibnamefont{Martinez}},
  \bibinfo{author}{\bibfnamefont{M.}~\bibnamefont{Robbiati}}, \bibnamefont{and}
  \bibinfo{author}{\bibfnamefont{S.}~\bibnamefont{Carrazza}},
  \emph{\bibinfo{title}{{QiNNtegrate}}} (\bibinfo{year}{2023}),
  \urlprefix\url{https://github.com/qiboteam/QiNNtegrate}.

\end{thebibliography}

\end{document}